\documentclass[aps, prx, notitlepage, citeautoscript, superscriptaddress, eqsecnum, reprint]{revtex4-2}

\usepackage{xr}
\usepackage{amsmath}
\usepackage{amsthm}
\usepackage{amsfonts, amssymb,amsxtra}
\usepackage[]{graphicx}
\usepackage{grffile}
\usepackage{amsfonts}
\usepackage{framed}
\usepackage{bbm}
\usepackage{braket}
\usepackage{xcolor}
\usepackage{mathtools}
\usepackage{LatexCommands}
\usepackage{booktabs}

\usepackage[colorlinks=true]{hyperref}
\hypersetup{
    unicode=false,          
    pdftoolbar=true,        
    pdfmenubar=true,        
    pdffitwindow=false,     
    pdfstartview={FitH},    
    pdftitle={},    
    pdfauthor={},     
    pdfsubject={},   
    pdfcreator={},   
    pdfproducer={}, 
    pdfkeywords={} {} {}, 
    pdfnewwindow=true,      
    colorlinks=true,       
    linkcolor=blue, 
    citecolor=blue,        
    filecolor=blue,      
    urlcolor=blue           
}

\begin{document}

\title{
Minimally Entangled Typical Thermal States for Classical and Quantum Simulation of 1+1-Dimensional $\mathbb Z_2$ Lattice Gauge Theory at Finite Temperature and Density}

\author{I-Chi Chen}
\thanks{These authors contributed equally to this work.}
\affiliation{Department of Physics and Astronomy, Iowa State University, Ames, Iowa 50011, USA}
\affiliation{Ames National Laboratory, U.S. Department of Energy, Ames, Iowa 50011, USA}

\author{Jo\~ao C. Getelina}
\thanks{These authors contributed equally to this work.}
\affiliation{Ames National Laboratory, U.S. Department of Energy, Ames, Iowa 50011, USA}

\author{Kl\'ee Pollock}
\affiliation{Department of Physics and Astronomy, Iowa State University, Ames, Iowa 50011, USA}

\author{Aleksei Khindanov}
\affiliation{Ames National Laboratory, U.S. Department of Energy, Ames, Iowa 50011, USA}

\author{Srimoyee Sen}
\affiliation{Department of Physics and Astronomy, Iowa State University, Ames, Iowa 50011, USA}

\author{Yong-Xin Yao}
\email{ykent@iastate.edu}
\affiliation{Department of Physics and Astronomy, Iowa State University, Ames, Iowa 50011, USA}
\affiliation{Ames National Laboratory, U.S. Department of Energy, Ames, Iowa 50011, USA}

\author{Thomas Iadecola}
\email{iadecola@iastate.edu}
\affiliation{Department of Physics and Astronomy, Iowa State University, Ames, Iowa 50011, USA}
\affiliation{Ames National Laboratory, U.S. Department of Energy, Ames, Iowa 50011, USA}

\begin{abstract}
Simulating strongly coupled gauge theories at finite temperature and density is a longstanding
challenge in nuclear and high-energy physics that also has fundamental implications for condensed
matter physics. 
In this work, we use minimally entangled typical thermal state (METTS) approaches to facilitate both classical and quantum computational studies of such systems. METTS techniques combine classical random sampling with imaginary time evolution, which can be performed on either a classical or a quantum computer, to estimate thermal averages of observables. 
We study 1+1-dimensional $\mathbb{Z}_2$ gauge theory coupled to spinless fermionic matter, which maps onto a local quantum spin chain. 
We benchmark both a classical matrix-product-state implementation of METTS and a recently proposed adaptive variational approach that is a promising candidate for implementation on near-term quantum devices, focusing on the equation of state as well as on various measures of fermion confinement. 
Of particular importance is the choice of basis for obtaining new METTS samples, which impacts both the classical sampling complexity (a key factor in both classical and quantum simulation applications) and complexity of circuits used in the quantum computing approach. 
Our work sets the stage for future studies of strongly coupled gauge theories with both classical and quantum hardware.
\end{abstract}
\date{\today}

\maketitle

\section{Introduction}
\label{sec:intro}
Gauge theories are archetypal models of strongly correlated matter that are relevant across energy scales.
In nuclear and high energy physics, the phase diagram of quantum chromodynamics (QCD) at finite density and temperature is of interest to the physics of the early universe, heavy ion collisions and neutron stars~\cite{Stephanov:2006zvm,Sorensen:2023zkk,Rajagopal:1999cp}.
In condensed matter physics, gauge theories play an important role in the description of topological phases of matter~\cite{wen2004quantum}, quantum criticality~\cite{Senthil2004,Grover16}, and correlated phenomena such as magnetism~\cite{Wegner71,Kogut79} and superconductivity~\cite{Senthil00,Sedgewick02,Lee07,Sachdev16}.
These diverse models share the key feature that they are generically strongly coupled such that their properties are beyond the reach of analytical tools.
For example, the outer and inner cores of neutron stars contain degenerate Fermi liquids of nucleons and/or quarks at densities close to few times the nuclear saturation density, where nuclear effective field theory and perturbative QCD calculations break down. 
Similar obstacles exist for calculations at higher temperature at finite density. 

Numerical tools to simulate gauge theories include Monte Carlo~\cite{Wilson74,Ceperley77,Blankenbecler81,Hirsch82,Foulkes01} and tensor network techniques~\cite{Schollwock11,Orus2014}, but both classes of methods have their limitations.
On one hand, Monte Carlo methods generically struggle to simulate systems at finite fermion density owing to the sign problem~\cite{Troyer05}.
On the other hand, tensor network simulations become much more challenging above one spatial dimension due to the complexity of contraction~\cite{Schuch07,Haferkamp20}.
Quantum simulation approaches bypass these obstacles and can potentially enable detailed studies of the phase diagram of strongly coupled gauge theories~\cite{Banuls20}.
However, much work remains to determine efficient quantum simulation methods for complex non-Abelian gauge theories like QCD~\cite{Bauer23}.
In the meantime, it is instructive to focus on method development for toy models that exhibit some of the same phenomenology as QCD, while still being relevant for condensed matter physics~\cite{Bernien17,Surace19,Yang20}.

In this paper we consider such a model, namely a $\mathbb Z_2$ lattice gauge theory coupled to spinless fermions in 1+1 dimensions~\cite{Borla20,Iadecola20,Yang20,Mildenberger22,Davoudi23,Desaules24}. 
Like QCD, this model exhibits chiral symmetry breaking, confinement and string tension~\cite{Borla20,Davoudi23,Kebrič24}; it is also relevant for studies of nonequilibrium condensed-matter phenomena like quantum many-body scars~\cite{Iadecola20,Mark20a,Aramthottil22,Gustafson23,Desaules24} and Hilbert-space fragmentation~\cite{Yang20}.
We explore the finite-temperature and -density properties of this model, focusing on measures of confinement and on the equation of state relating internal energy and fermion density.
Our study adopts the minimally entangled typical thermal states (METTS) approach~\cite{White09,stoudenmire2010,Binder15,Binder17,Wietek21a,Wietek21b}, which combines imaginary time evolution (ITE) with a statistical sampling procedure to estimate quantum statistical-mechanics averages.
Although originally developed as a tensor-network method, METTS can also be recast as a quantum algorithm (QMETTS), in which a quantum computer is used to perform the ITE subroutine~\cite{Motta19}.
While many approaches to ITE on quantum computers are possible, our study focuses on a recently proposed adaptive~\cite{AVQITE} variational~\cite{McArdle19} approach to QMETTS (AVQMETTS)~\cite{Getelina23}.
We perform systematic benchmarks of grand-canonical-ensemble calculations within METTS with a view towards both classical and quantum computing approaches, focusing on a matrix product state (MPS) approach for the former and on exact statevector calculations for the latter.
The classical and quantum approaches have common systematics in that they rely on the same classical sampling procedure, where the choice of sampling basis has an impact on convergence.
AVQMETTS additionally depends on a choice of the operator pool used to construct the variational ansatz state, which impacts both the accuracy and quantum resource cost of the simulation.
In carefully examining these systematics, our study lays groundwork for future progress in both classical and quantum simulation of gauge theories.

The remainder of the paper is organized as follows. 
In Sec.~\ref{sec:model}, we define the $\mathbb Z_2$ gauge theory model and review how to perform grand-canonical-ensemble calculations within METTS.
In Sec.~\ref{sec:metts}, we benchmark classical METTS calculations of the internal energy density $\epsilon$ and fermion density $n$, focusing in particular on how to choose the optimal sampling basis.
We then present METTS calculations of the $\epsilon$-$n$ equation of state, as well as two measures of confinement: Friedel oscillations of the fermion density~\cite{Borla20} and string-antistring distribution functions~\cite{Kebrič24}.
In Sec.~\ref{sec:avqmetts}, we review the AVQMETTS method before benchmarking its performance with respect to sampling basis and operator pool.
We move on to show results for the equation of state and Friedel oscillations, which we argue are a robust probe of confinement for the relatively small system sizes accessible on today's quantum computers.
In addition, we perform statevector and tensor-network simulations of the AVQMETTS algorithm and study the system-size scaling of its circuit complexity up to a system size of $L=32$, observing a power-law dependence.
We end in Sec.~\ref{sec:outlook} with an outlook for future work.

\section{$\mathbb Z_2$ Lattice Gauge Theory and METTS}\label{sec:model}
We consider a $1+1$-dimensional model of spinless and massless fermions coupled to a $\mathbb Z_2$ gauge field:
\begin{align}
\label{eq:z2lgt}
    H=\frac{1}{2}\sum_{i=1}^{L-1}\left(c^{\dagger}_{i}\sigma^{z}_{i,i+1}c_{i+1}+\text{H.c.}\right)+h\sum_{i=0}^{L}\sigma^{x}_{i,i+1}\,,
\end{align}
where the first term is the kinetic term, and the second represents the confining (electric) field with strength $h$.
The fermions are represented by creation/annihilation operators $c^\dagger_i$/$c_i$ on site $i=1,\dots,L$ and the $\mathbb Z_2$ gauge field is represented by Pauli operators $\sigma^z_{i,i+1}$ and $\sigma^x_{i,i+1}$ on the links $(i,i+1)$.
Note that the 1D lattice in Eq.~\eqref{eq:z2lgt} contains $L$ fermion sites and $L+1$ gauge links; the states of the gauge links $(0,1)$ and $(L,L+1)$ are constants of the motion because no fermion can hop across those links (we assume open boundary conditions).
The Hamiltonian \eqref{eq:z2lgt} commutes with the Gauss-law operators $G_{i}=\sigma_{i-1,i}^{x}\left(-1\right)^{n_{i}}\sigma_{i,i+1}^{x}$ ($i=0,\dots,L$) with $n_i=c_{i}^{\dagger}c_{i}$ the fermion number density. 
Each selection of $\langle G_i\rangle =\pm 1$ (known as a background charge configuration) corresponds to an independent symmetry sector of the Hamiltonian. 
In this paper we choose the uniform background charge configuration $\langle G_i\rangle=1$.
The model \eqref{eq:z2lgt} can be recast as a pure spin-1/2 model in terms of gauge-invariant local operators $Z_{i}=\sigma_{i,i+1}^{x}$ ($i=0,\dots,L$) and $X_{i}=(c^\dagger_i-c_i)\sigma_{i,i+1}^{z}(c^\dagger_{i+1}+c_{i+1})$ ($i=1,\dots,L-1$) as follows \cite{Borla20}:
\begin{align}
    H=\frac{1}{4}\sum_{i=1}^{L-1}\left(X_{i}-Z_{i-1}X_{i}Z_{i+1}\right)+h\sum_{i=0}^{L}Z_{i}. 
\label{eq:hamiltonian}
\end{align}
Note that the operators $Z_0$ and $Z_L$ are not dynamical (i.e. the operators $X_0$ and $X_L$ are undefined) and therefore serve only to label the states of the frozen gauge links $(0,1)$ and $(L,L+1)$.

To study finite-temperature and -density properties of the model~\eqref{eq:hamiltonian}, we compute the thermal expectation value of a generic observable $\mathcal{O}$ in the grand canonical ensemble at inverse temperature $\beta=1/T$ and chemical potential $\mu$: 
\begin{equation}
    \left\langle \mathcal{O}\right\rangle _{\mu,\beta}=\frac{1}{\mathcal{Z}} \text{Tr}\left(\mathcal{O}e^{-\beta(H-\mu N)}\right)\,,
    \label{eq:th_val}
\end{equation}
where $N$ is the total fermion number operator and $\mathcal{Z} = \text{Tr}\left(e^{-\beta(H-\mu N)}\right)$ is the grand canonical partition function. 
In the spin model, the fermion number operator is given by
\begin{align}
\begin{split}
    N&=\sum_{i=1}^{L}n_i\,, \label{eq:number_op} \\
    n_i&=\frac{I-Z_{i-1}Z_{i}}{2}\,,
\end{split}
\end{align}
which corresponds to the total number of Ising domain walls. 
To evaluate Eq.~\eqref{eq:th_val}, we adopt a statistical sampling procedure defined by the METTS algorithm.

A METTS calculation can be viewed as a Markovian random walk consisting of multiple ``thermal steps." The first thermal step starts with a random classical product state (CPS) $\ket{i}$ and performs ITE up to an imaginary time $\tau=\beta/2$ to obtain a METTS, which written as 
\begin{equation}
\ket{\phi_{i}\left(\beta\right)} =P_{i,\beta}^{-1/2}e^{- (\beta/2) (H-\mu N)}\left|i\right\rangle \,,
\end{equation}
where $P_{i,\beta}=\left\langle i\left|e^{-\beta (H-\mu N)}\right|i\right\rangle $ is a normalization factor. 
We henceforth drop the explicit dependence on $\beta$ to simplify the notation.
A sample of the thermal average of an observable $\mathcal{O}$ is then given by $\left\langle \mathcal{O}\right\rangle _{i}=\left\langle \phi_{i}\left|\mathcal{O}\right|\phi_{i}\right\rangle$. 
The next thermal step is triggered by an all-qubit measurement collapse of the METTS $\ket{\phi_i}$ to a CPS $\ket{i'}$ in a specific basis, which occurs with probability $\left|\left\langle i'|\phi_{i}\right\rangle \right|^{2}$.
Thus, one thermal step amounts to a transition between CPSs $\ket{i}$ and $\ket{i'}$ in a Markovian random walk.
The stationary distribution of this process is simply $P_i/\mathcal Z$, owing to the detailed balance condition $\left|\left\langle i'|\phi_{i}\right\rangle \right|^{2}/\left|\left\langle i|\phi_{i'}\right\rangle \right|^{2} = P_{i'}/P_{i}$.
For an ensemble obtained from $S$ thermal steps, the thermal expectation value of an observable $\mathcal{O}$ can then be estimated as:
\begin{equation}
\left\langle \mathcal{O}\right\rangle _{\text{METTS}}=\frac{1}{S}\sum_{i=1}^{S}\left\langle \mathcal{O}\right\rangle _{i} \,.
\end{equation}
In practice, the METTS sampling is performed in parallel with $S_\textrm{w}$ independent random walks, each of which generates $S_0$ METTSs. 
This amounts to an ensemble of size $S=S_\textrm{w}S_0$. 
To remove memory of the initial conditions, the first few thermal steps (typically the first ten) are excluded from the statistical analysis. 
Furthermore, the choice of measurement basis for METTS collapse is flexible, and previous investigations have demonstrated that alternating between $x$- and $z$-basis measurements (abbreviated as $xz$-basis) between adjacent thermal steps drastically reduces autocorrelations of the random walk~\cite{stoudenmire2010,Binder17,Getelina23}.
Here we perform a more systematic study of the choice of collapse basis and adopt the best strategy for our specific applications.

ITE is the key subroutine in the METTS sampling approach. 
For gapped systems in 1D, the standard classical algorithm uses tensor networks (specifically MPSs) owing to the fact that ground states of such systems have finite bipartite entanglement and therefore typically feature a system-size-independent bond dimension $\chi$.
However, the scaling for tensor network approaches becomes less favorable for critical systems or above one spatial dimension, motivating quantum computing approaches as discussed in Sec.~\ref{sec:intro}.
We therefore additionally benchmark the AVQITE algorithm~\cite{AVQITE} for METTS preparation, presented previously as the AVQMETTS approach~\cite{Getelina23}. 
AVQITE features automatically generated compact circuits for state propagation, amenable to near-term quantum computing.
The MPS calculations discussed in this work were performed using the iTensor package~\cite{itensor}, while the AVQMETTS calculations were performed using the AVQITE code~\cite{CyQC}.

\section{Classical METTS calculations}\label{sec:metts}
\begin{figure*}
     \includegraphics[width=0.9\textwidth]{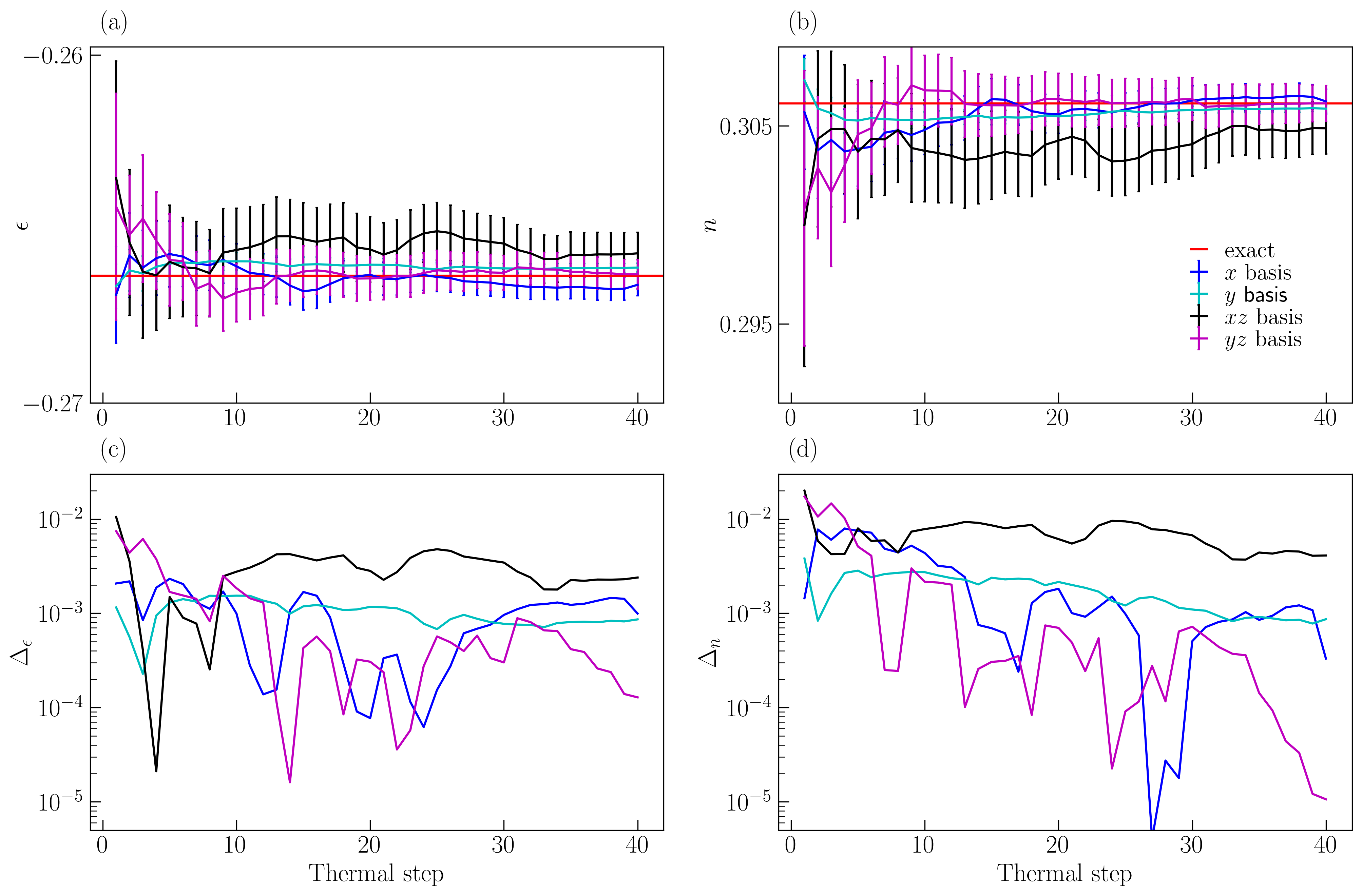}
     \caption{Classical METTS calculations of energy and particle density in the model \eqref{eq:hamiltonian} with $L=12$ at $\beta=10$ using various choices of measurement basis for state collapse. We set $h=0.1$ and $\mu= -0.4$ to reach a filling of $n=\frac{1}{3}$ in the ground state. (a) Estimated energy density $\epsilon = \frac{1}{L}\frac{1}{S_\text{w} k} \sum_{j=1}^{k}\sum_{i=1}^{S_\text{w}} \langle \phi_{ij}|\,H\,|\phi_{ij}\rangle$ as a function of thermal step $k$ with $S_\text{w}=100$ parallel random walks. The choices of measurement basis include $y$ (cyan line), $x$ (blue line), and alternating $xz$ (black line) and $yz$ bases (magenta line). For reference, the energy density obtained with exact diagonalization (ED) is also shown (red line). The error bars are plotted as the standard errors of $\epsilon$. 
     (b) Same as (a) but for the particle density $n$.
     (c) Relative error $\Delta_\epsilon$ of the energy density $\epsilon$ with increasing number of thermal steps. (d) Relative error $\Delta_n$ of the particle density $n$. The relative error is defined as $\Delta_\mathcal{O} = \frac{\left| \left\langle \mathcal{O} \right\rangle_\textrm{METTS} - \left\langle \mathcal{O} \right\rangle_\textrm{ED} \right|}{\left| \left\langle \mathcal{O} \right\rangle_\textrm{ED} \right|}$ for a generic observable $\mathcal{O}$.
     }
     \label{fig:E_n_N}
 \end{figure*}

Before discussing the AVQMETTS simulations, we perform finite-temperature classical METTS calculations for the model~\eqref{eq:hamiltonian} leveraging the efficiency of MPSs. 
The purpose is two-fold. 
First, as noted in Sec.~\ref{sec:model}, the choice of METTS sampling basis can have a strong effect on sampling efficiency---for example the $xz$-basis collapse can outperform $z$-basis collapse as observed in Refs.~\cite{stoudenmire2010,Getelina23}.
Moreover, in Ref.~\cite{Chen23} it was observed in a different context that sampling infinite-temperature expectation values in the $y$-basis can be advantageous for models like Eq.~\eqref{eq:hamiltonian} whose Hamiltonians contain only Pauli-$X$ and $Z$ terms.
Thus we aim to study more systematically the comparative performance of collapse in various bases, including the $y$ and $x$-basis, as well as alternating $yz$ basis and $xz$ collapse bases along the lines discussed in Sec.~\ref{sec:model}. 
We will use the optimal basis choice for subsequent classical METTS calculations and contrast the strategy for AVQMETTS, where additional constraints, such as circuit complexity measured by number of two-qubit gates, have to be taken into consideration. 
Second, we aim to have numerically converged calculations of the equation of state of the model \eqref{eq:hamiltonian}, utilizing the efficient Trotter approach for ITE in the MPS basis. 
Specifically, we evaluate the energy density $\epsilon =\left\langle H\right\rangle_{\mu,\beta} /L$ and particle density $n =\left\langle N\right\rangle_{\mu,\beta} /L$ as functions of temperature and chemical potential.
These calculations elucidate the physics of the 1D model and provide benchmark data for the AVQMETTS simulations presented in Sec.~\ref{sec:avqmetts}.

\subsection{Optimal measurement basis for METTS collapse}

Fig.~\ref{fig:E_n_N} visualizes the estimated energy density $\epsilon$ and particle density $n$ for $L=12$ with $\beta=10$ and $\mu=-0.4$ as a function of the number of thermal steps, highlighting the dependence of convergence on several specific choices of measurement basis for METTS collapse. 
These data are obtained using $S_\text{w} = 100$ parallel random walks. 
Figure~\ref{fig:E_n_N}(a) shows that $\epsilon$ obtained from METTS calculations with four different basis choices are generally quite close to the exact reference energy density, with relative error $\Delta_{\epsilon} \lesssim 1\%$ as shown in Fig.~\ref{fig:E_n_N}(c). 
For larger numbers of thermal steps, $\Delta_{\epsilon}$ is consistently smaller for calculations using the $x$-, $y$- or $yz$-bases than that using the $xz$-basis. 
The fluctuations of $\Delta_\epsilon$ with thermal steps are consistent with the standard error of $\epsilon$. 
The $y$-basis calculation gives the smallest standard error ($\gtrsim 0.6\times 10^{-3}$), which aligns with the minimal fluctuations of $\Delta_\epsilon$.
This is consistent with the findings of Ref.~\cite{Chen23} which observed that the $y$-basis displayed minimal shot-to-shot fluctuations when sampling energy-density correlation functions at infinite temperature.
Similar observations apply to the calculations of particle density $n$, with slightly larger errors occuring for the $xz$-basis results, as shown in Fig.~\ref{fig:E_n_N}{(b,d)}. 
In the following classical METTS simulations we choose the alternating $yz$-basis for state collapse, since calculations using this basis give consistently lower errors at large numbers of thermal steps. 

\subsection{Equation of state} \label{sec: eos}
\begin{figure}[t]
     \includegraphics[width=0.49\textwidth]{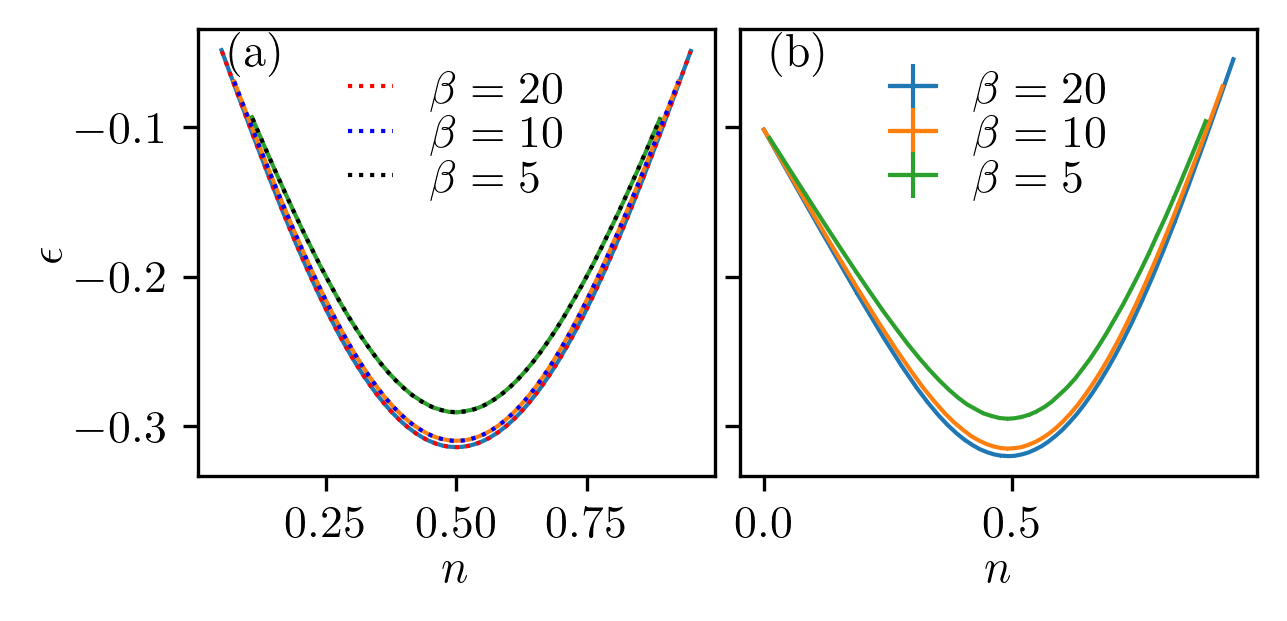}
     \caption{Energy density $\epsilon$ versus particle density $n$ equation of state obtained from METTS calculations (solid lines) on the model \eqref{eq:hamiltonian} for $L=60$ in the deconfined free-fermion limit $h=0$ (a) and in the confined phase with $h=0.1$ (b). The curves are shown for three inverse temperatures $\beta=5, 10$ and $20$. The analytical results for the free fermion model are also plotted in (a) for reference as dotted lines, which agree perfectly with the respective METTS calculations. 
     }
     \label{fig:E_n}
 \end{figure}

To study the equation of state for the model~\eqref{eq:hamiltonian}---\textit{i.e.}, the functional relationship between energy density $\epsilon$ and particle density $n$---we perform METTS calculations of a $L=60$ model with chemical potential varying from $\mu=-1.0$ to $\mu=1.0$ with a step $0.025$. 
The results for the free-fermion limit $h=0$ are plotted in Fig.~\ref{fig:E_n}(a) at three inverse temperatures $\beta=5, 10$ and $20$. 
Without the confining field, the energy-particle density curve is symmetric around half-filling, indicative of particle-hole symmetry.
Generally, the thermal energy density increases with temperature, as the number of excited states within the accessible energy window $\sim 1/\beta$ increases. 
As a result, $\epsilon$ grows the most at half-filling, and becomes trivially temperature-independent at zero or full filling, where the relevant subspace dimension reduces to 2 and the ground state is two-fold degenerate. 
For reference, the analytical calculations for the same free-fermion model are also shown in Fig.~\ref{fig:E_n}(a) with dotted lines, which agree perfectly with the METTS results and confirm the convergence of METTS calculations.
In Fig.~\ref{fig:E_n}(b), we show results for the equation of state with $h=0.1$. 
The finite confining field clearly breaks the particle-hole symmetry. 
Compared with the free-fermion results in Fig.~\ref{fig:E_n}(a), the energy density becomes significantly lower for $n < 0.5$ and reaches maximal energy reduction at $n = 0$.
In contrast, the variation of $\epsilon$ for $n > 0.5$ is much smaller, and remains the same at $n = 1$. 
This can be understood by considering the $\mathbb{Z}_2$ symmetry breaking by the confining field, which promotes the configurations with large negative magnetizations (or equivalently, those with long anti-strings as defined in Sec.~\ref{sec: FO}) and further lowers the energy. 
Since the total magnetization magnitude (length of anti-string) in each particle number sector is bounded by $L-2\lfloor N/2\rfloor$, the impact of the confining field on the thermal statistics grows from large filling to small filling of the model, consistent with the curves shown in Fig.~\ref{fig:E_n}(b).
Note that the standard errors of $\epsilon$ and $n$ for these METTS calculations are below $10^{-3}$, \textit{i.e.} they are are smaller than the line width of the curves.
Furthermore, we have checked by performing additional METTS calculations at $L=120$ that increasing the system size results in an almost imperceptible change of the equation of state curves shown in Fig.~\ref{fig:E_n}(b).
This builds confidence that these results are already close to the thermodynamic limit for $L\lesssim60$.


\subsection{Probes of confinement: Friedel oscillations and string length distributions} \label{sec: FO}

 \begin{figure}[t]
     \includegraphics[width=0.475\textwidth]{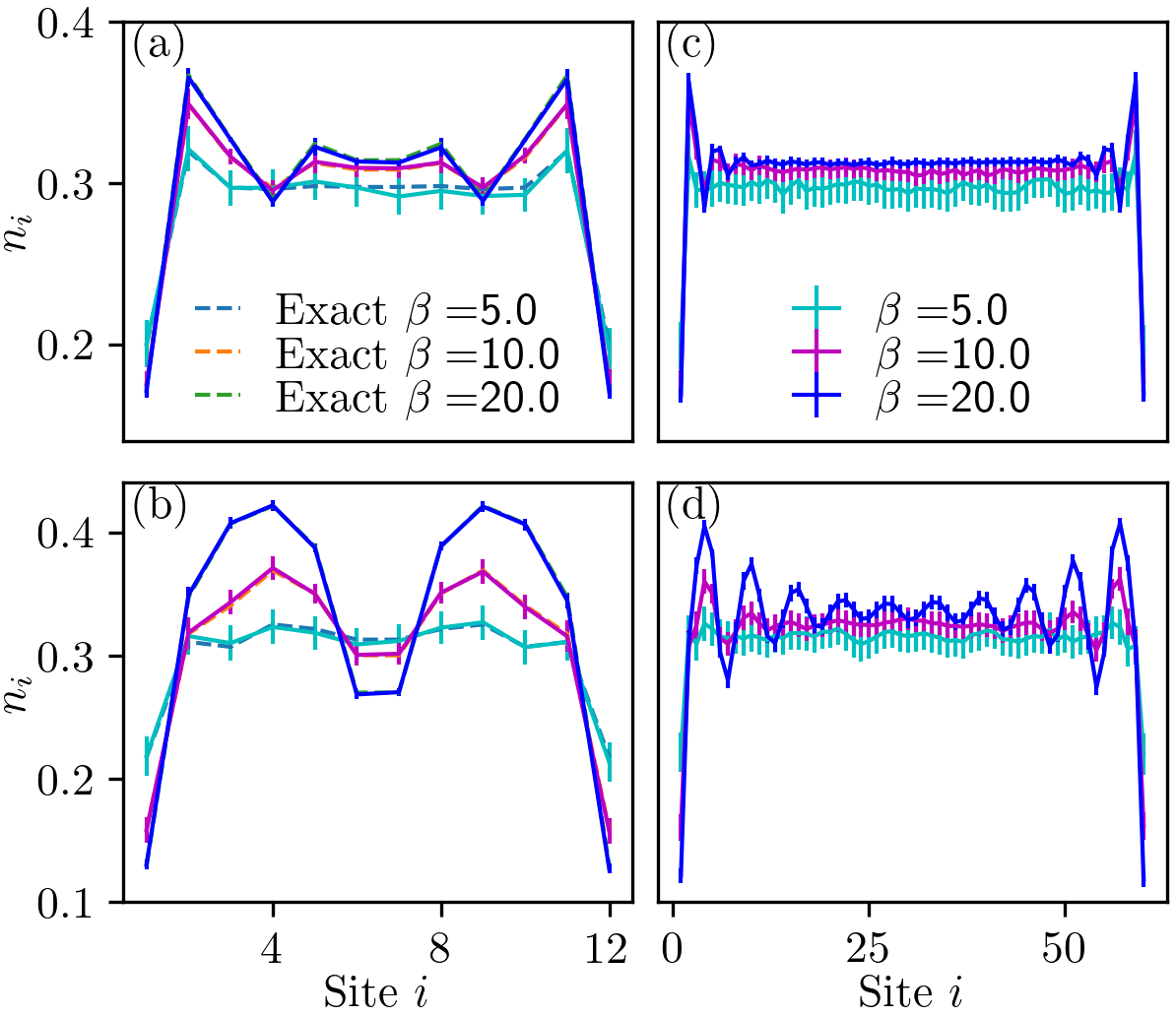}
     \caption{Friedel oscillations and their dependence on confining field and temperature. (a) Spatial distributions of single site fermion occupation numbers $\braket{n_i}$ (solid lines) for $L=12$ and $h=0$ at three representative temperatures $\beta=5, 10$ and $20$. (b) Same as (a) but with confining field $h=0.1$. (c) Same as (a) but for $L=60$. (d) Same as (a) but for $L=60$ and $h=0.1$. 
     We use $S_\text{w} = 100$ random walks to generate $S_\text{w}\times S_0 = 2000$ METTS samples for the analysis. The particle occupation distributions obtained from exact diagonalization at $L=12$ are also shown as dashed lines in (a,b). The particle filling is set to $\frac{1}{3}$ per site at zero temperature by introducing a chemical potential $\mu=-0.55$ for $h=0$ and $\mu=-0.4$ for $h=0.1$.
     }
     \label{fig:fo}
 \end{figure}

\begin{figure*}
     \includegraphics[width=\textwidth]{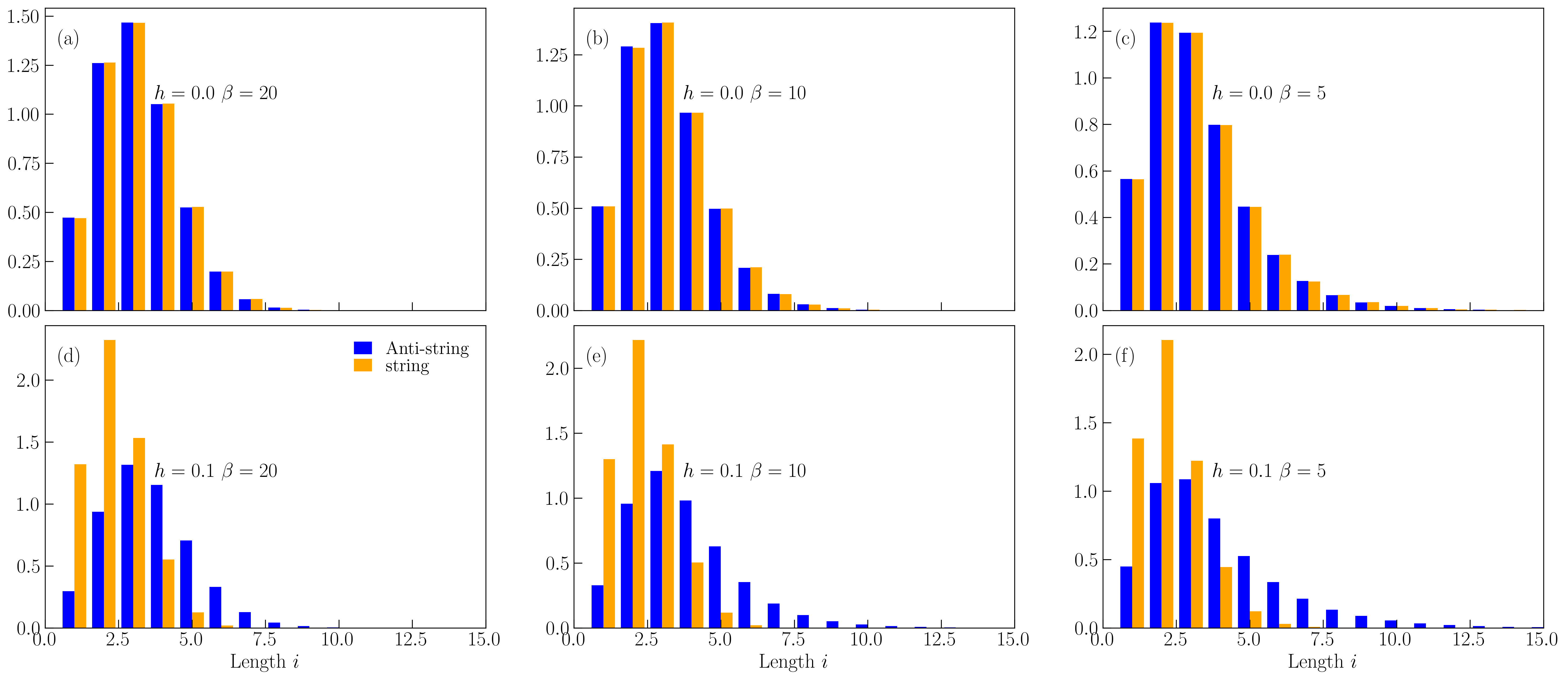}
     \caption{(Anti-)string length distribution $C_l$ as a function of temperature and confining field. The (anti-)string length distributions are plotted as (blue) orange histograms with $h=0$ at $\beta=20$ (a), $\beta=10$ (b), $\beta=5$ (c) and $h=0.1$ at $\beta=20$ (d), $\beta=10$ (e), $\beta=5$ (f). Calculations are performed with $L=36$ at fixed ground-state particle filling $n=1/3$. Here $C_l$ is evaluated for each (anti-)string length $l$ as the number of counts for that $l$ divided by the total number ($10^5$) of bit-string samples. Note that individual $C_l$s can be greater than one because there can be multiple (anti-)strings of the same length in the same bit-string.
     }
     \label{fig:Strl}
 \end{figure*}

A fundamental feature of the $\mathbb{Z}_2$ lattice gauge theory~\eqref{eq:z2lgt} is its manifestation of fermion confinement in both the ground state~\cite{Borla20} and the dynamics~\cite{Yang20}.
For example, in the ground state at finite $h$, fermions form tightly bound pairs known as mesons owing to the fact that the confining field $h$ imposes an energy cost scaling linearly with the separation between fermions.
This behavior is also evident in the equivalent spin model \eqref{eq:hamiltonian}, where $h$ effectively counts the number of flipped spins between two Ising domain walls.
One way to characterize confinement in this model is to examine Friedel oscillations in the spatial particle density distribution at specific particle fillings~\cite{Borla20}. 
At zero temperature and $h=0$ the wavelength of the Friedel oscillation is simply $1/2k_F$, where $k_{\rm F}$ is the Fermi wavevector. 
In contrast, with a finite confining field, each pair of particles forms a bound state so that the wavevector reduces to $k_\text{F}$, leading to a doubling of the wavelength.
Here we perform METTS calculations to investigate the temperature dependence of these Friedel oscillations.
These calculations are complementary to simulations performed using purification approaches~\cite{Kebrič24} and provide further benchmark results for AVQMETTS calculations.
We set the particle filling $n = 1/3$ in the ground state by tuning the chemical potential to $\mu=-0.55$ for the free fermion model and $\mu=-0.4$ with confining field $h=0.1$.

In Fig.~\ref{fig:fo}, we demonstrate the Friedel oscillations by plotting the spatial distribution of site-wise particle occupations $\braket{n_i}$ for various temperatures and two values of the confining field.
Figure~\ref{fig:fo}(a) shows that there are four peaks in the spatial distribution of $\braket{n_i}$ at $\beta=20$ for $L=12$ and $h=0$, which corresponds to $n = 1/3\times 12 = 4$ particles in the system. 
As temperature increases, the amplitude of these peaks gradually diminishes. 
By $\beta=5$, the two central peaks disappear.
Results from calculations at finite confining field $h=0.1$ are shown in Fig.~\ref{fig:fo}(b), where the number of peaks in the particle density distribution reduces to 2, as expected from the meson formation mechanism. 
Temperature plays a similar role and gradually washes away the peaks, suggesting a temperature-induced crossover between confined and deconfined regimes. 
Similar observations apply to larger system sizes readily accessible within METTS.
For the $L=60$ free fermion model ($h=0$), while the Friedel oscillations near edges are visible, their amplitude is damped towards the middle of the chain,
as shown in Fig.~\ref{fig:fo}(c). 
Interestingly, the Friedel oscillations are enhanced with finite $h=0.1$, as depicted in Fig.~\ref{fig:fo}(d). 
The ten visible peaks correspond to $n = 1/3\times 60/2 = 10$ mesonic bound states, and the peak damps two times slower 
toward the center of the chain.

Another feature distinguishing the confined and deconfined regimes is the disparity in string and anti-string length distributions~\cite{Kebrič24}. 
Here an (anti-)string refers to the consecutive 1s (0s) in a bitstring obtained from measuring a thermal state in the $z$-basis of the spin model \eqref{eq:hamiltonian}; the endpoints of these (anti-)strings are Ising domain walls, which correspond to the fermions in Eq.~\eqref{eq:z2lgt}. 
In the deconfined limit $h=0$, the length distributions are symmetric between strings and anti-strings, which reflects the $\mathbb{Z}_2$ symmetry in the absence of a confining field. 
Strings of shorter lengths become more pronounced in the confined regime.
Here the confining field raises the energy cost of configurations with longer strings while lowering the energy for those with longer anti-strings, leading to the formation of mesonic bound states.
In practice, to get the (anti-)string length statistics, we first collapse each of the $S_\text{w}\times S_0=2000$ METTSs $50$ times in the $z$-basis, to obtain in total $2000\times 50$ bit-strings. We then count the number of consecutive 1s (0s) in each bit-string to obtain the (anti-)string lengths.
For example, in the bit-string $100111$ we would record one occurrence each for strings of length 1 and 3 and one anti-string of length 2.

Figure~\ref{fig:Strl} shows the (anti-)string length distributions for an $L=36$ model at particle filling $n = 1/3$ with and without confining field. 
In Fig.~\ref{fig:Strl}(a) we see that the length distributions of strings and anti-strings at $\beta=20$ are nearly identical, with a skewed distribution and a peak at $l=3$ tied to the $1/3$ particle filling. 
Upon increasing temperature to $\beta=10$ (Fig.~\ref{fig:Strl}(b)) and $\beta=5$ (Fig.~\ref{fig:Strl}(c)), the symmetric (anti-)string length distributions remain but are broadened due to thermal smearing effects. 
The string and anti-string length distributions exhibit opposite trends upon turning on the confining field $h=0.1$ as shown in Fig.~\ref{fig:Strl}(d).
While the center of the anti-string distribution shifts to larger length $l$, the string distribution shifts to smaller $l$, consistent with the confinement effect discussed above. 
Increasing temperature similarly broadens the distributions as demonstrated at $\beta=10$ [Fig.~\ref{fig:Strl}(e)] and $\beta=5$ [Fig.~\ref{fig:Strl}(f)], indicating a tendency toward deconfinement.

Our results on Friedel oscillations and (anti-)string statistics at finite temperature are consistent with those obtained in Ref.~\cite{Kebrič24} with a complementary MPS approach based on purifications~\cite{Zwolak04,Verstraete04,Feiguin05}.
The purification method computes thermal averages by preparing a state in which each physical qubit is maximally entangled with an auxiliary qubit, then performing ITE on the physical qubits for an imaginary time $\beta/2$.
This method avoids the sampling entailed in METTS at the expense of doubling the number of qubits, which limits the system sizes that are easily accessible.
The METTS approach adopted here allows us to simulate chains of up to (and even beyond) $L=60$ fermion sites as in Figs.~\ref{fig:E_n} and \ref{fig:fo}, whereas the finite-temperature calculations in Ref.~\cite{Kebrič24} were limited to systems of $\sim36$ sites.
This increase in the accessible system sizes may be useful when studying, e.g., finite-temperature confinement phase diagrams in higher dimensions, where reducing the qubit overhead is particularly important.

\section{AVQMETTS simulations}
\label{sec:avqmetts}

In this section, we turn to a quantum version of METTS, specifically the AVQMETTS approach~\cite{Getelina23}, for finite-temperature simulations of the model~\eqref{eq:hamiltonian}. 
Here we focus on benchmarking AVQMETTS calculations of the 1D model, motivated by the intriguing physics and available exact solutions. 
This section is divided into three parts. We first briefly outline AVQMETTS method and its technical details. 
Secondly, we present our AVQMETTS numerical results concerning the finite-temperature behavior of the energy and particle densities. 
Finally, we give a numerical estimation of the system-size scaling for the AVQMETTS circuit complexity characterized by the number of CNOT gates, which is a useful resource estimate for near-term applications.
Even though AVQMETTS is a quantum algorithm, all the following calculations are performed using classical simulations, which provide important benchmarks for future AVQMETTS calculations on quantum processing units.

\subsection{The AVQMETTS method}

The AVQMETTS approach adopts AVQITE~\cite{AVQITE}, rather than the standard MPS-based ITE algorithms such as Trotter decomposition or time-dependent variational principle~\cite{Haegeman11, Haegeman16}, for METTS preparation as outlined in Sec.~\ref{sec:model}. 
The AVQITE algorithm dynamically constructs a parameterized circuit $\left| \psi(\boldsymbol{\theta}[\tau]) \right\rangle = \prod_j e^{-\frac{1}{2}i\theta_j[\tau] G_j}\left| \psi_0 \right\rangle$ that closely follows the exact dynamics in imaginary time $\tau$ generated by the Hamiltonian $H$ for a CPS $\ket{\psi_0}$.
Here, $G_j$ are Hermitian generators drawn from a predetermined operator pool (see below).
The parameters evolve according to the equations of motion: 
\begin{equation}
    \sum_j g_{ij} \dot{\theta}_j = V_i\,,  \label{eq: eom}
\end{equation}
where $g_{ij} = \Re[\langle \frac{\partial \psi}{\partial\theta_i}|\frac{\partial \psi}{\partial\theta_j}\rangle + \langle \psi|\frac{\partial \psi}{\partial\theta_i}\rangle \langle \psi|\frac{\partial \psi}{\partial\theta_j}\rangle ]$, also known as the Fubini-Study metric, is the real part of the quantum geometric tensor, and where $V_i = -\Re[\langle \frac{\partial \psi}{\partial\theta_i}|H|\psi\rangle]$ is the energy gradient with respect to variational parameter $\theta_i$. 

The accuracy of the variational dynamics is maintained by monitoring the McLachlan distance, $L^2 = 2(\langle \psi |H^2|\psi \rangle - \langle \psi |H|\psi \rangle^2 -\sum_i V_i\dot{\theta}_i)$~\cite{Yuan19, AVQITE, Yao-AVQDS-PRX_Q-2021}, which measures the separation between the variationally and exactly propagated states. 
If the McLachlan distance surpasses a preset threshold (e.g., $10^{-3}$) at a certain time step, the ansatz is expanded by appending more parameterized unitaries $\mathcal{U}_j$ to it until the distance threshold is met.
The unitary $\mathcal{U}_j = e^{-\frac{1}{2}i\theta_j G_j}$ is obtained by selecting from the operator pool the generator $G_j$ which maximally reduces the McLachlan distance.
A variety of operator pools have been explored, and the choice depends on the problem under study and can affect the convergence of AVQITE calculations~\cite{AVQITE, Getelina24}.

Before continuing we provide a few technical details about our implementation of the AVQMETTS algorithm.
When propagating the variational parameters $\boldsymbol{\theta}[\tau] \to \boldsymbol{\theta}[\tau] + \boldsymbol{\dot{\theta}} \delta \tau$ according to Eq.~\eqref{eq: eom}, we adopt a variable time step $\delta \tau$ such that $\max_j |\dot{\theta_j}|\delta \tau$ is fixed~\cite{Yao-AVQDS-PRX_Q-2021}, as a slight improvement in accuracy is observed compared with AVQITE simulations with a comparable fixed time step $\delta \tau = 0.02$.
Furthermore, we consider three different operator pools depending on the measurement basis used to collapse the METTSs. For propagating an initial CPS in the $z$-basis, the operator pool is the simplest and given by $\mathcal{P}_z = \{Y_i\}_{i=0}^L \cup \{Y_iZ_j\}_{0\le i\neq j \le L}$. For a CPS in the $x$-basis, the operator pool is expanded to $\mathcal{P}_x = \mathcal{P}_z \cup \{Y_iX_j\}_{0\le i\neq j \le L} \cup \{Y_iZ_jX_k\}_{0\le i\neq j \neq k\le L}$. Finally, for a CPS in the $y$-basis, we adopt an operator pool given by $\mathcal{P}_y = \{Z_i,X_i\}_{i=0}^L \cup \{Z_i X_j\}_{0\le i\neq j \le L} \cup \{Z_i Z_j Z_k\}_{0\le i < j < k \le L}$. We constructed the $y$-basis operator pool by performing a preliminary round of calculations where we compared the fidelity between CPSs evolved under AVQITE and under exact ITE, and we selected those Pauli operators generating the highest fidelity. Note that multiple operator pools will generically be used in the course of an AVQMETTS calculation with a collapse basis that alternates between thermal steps as described in Sec.~\ref{sec:model}.

\begin{figure*}
     \includegraphics[width=\textwidth]{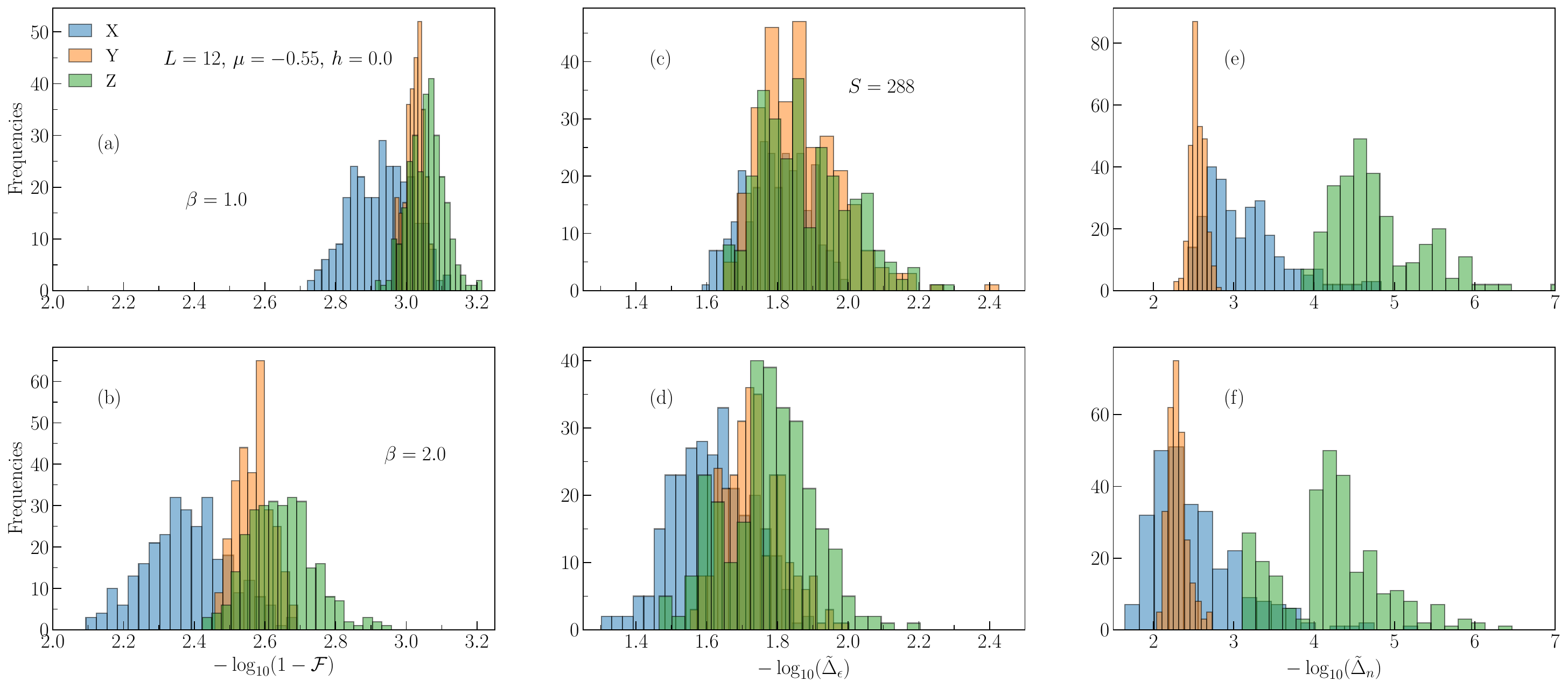}
     \caption{Benchmarking the accuracy of AVQITE calculations for initial CPSs in the $x$, $y$, and $z$-basis.
     (a) Histogram of the infidelities $1-\mathcal{F}$ at $\beta=1$. (b) Same as (a) but at $\beta=2$. 
     (c) Histogram of the relative errors of the energy $E=\langle H \rangle_\text{AV}$ at $\beta=1$. (d) the same as (c) but at $\beta=2$.
     (e) Histogram of the relative errors of the particle filling $\langle N \rangle_\text{AV}$ at $\beta=1$. (f) the same as (e) but at $\beta=2$.
     We consider a random sample of initial CPSs of size $\mathcal M=288$ in each basis.  The Hamiltonian parameters are set to $L=12$, $\mu=-0.55$, and $h=0$ to yield a ground state at $n=1/3$ filling.}
     \label{fig:plt_acc}
\end{figure*}

\subsection{Benchmarking the impact of CPS collapse basis}

As discussed in Sec.~\ref{sec:metts}, the choice of collapse basis plays an important role in the METTS sampling procedure.
However, this choice must be re-evaluated in the context of AVQMETTS, which uses AVQITE as the CPS propagation subroutine.
For example, the performance and quantum resource requirements of AVQITE simulations depend on the choice of reference state $\ket{\psi_0}$, which in the context of AVQMETTS calculations can be a product state in the $x$, $y$, or $z$ basis depending on the choice of collapse basis.
To explore this dependence, we benchmark the accuracy of AVQITE simulations of the model~\eqref{eq:hamiltonian} in comparison with ITE via exact diagonalization.
We consider a system of size $L=12$ in the free-fermion limit ($h=0$) and with a chemical potential of $\mu=-0.55$, such that the ground state has four particles (1/3 filling). 
We then randomly select $\mathcal{M}=288$ CPSs in each of the three ($x/y/z$) measurement bases, and evolve them using both AVQITE and exact ITE up to an imaginary time $\beta/2$.
We consider two figures of merit for the accuracy of AVQITE simulations relative to exact ITE.
First, we consider the infidelity $1-\mathcal{F}=1-\left|\left\langle \psi(\beta)_\textrm{AV} \right. \left| \psi(\beta)_\textrm{ITE}\right\rangle\right|^2$ of the AVQITE state $\left| \psi(\beta)_\textrm{AV}\right\rangle$ with respect to the exact ITE state $\left| \psi(\beta)_\textrm{ITE}\right\rangle$. 
Second, we consider the quantity $\tilde\Delta_\mathcal{O} = \left| \left\langle \mathcal{O} \right\rangle_\textrm{AV} - \left\langle \mathcal{O} \right\rangle_\textrm{ITE} \right|/\left| \left\langle \mathcal{O} \right\rangle_\textrm{ED} \right|$, where $\braket{O}_{\rm ED}$ is the thermal average of $\mathcal O$ obtained from exact diagonalization. 
This quantity can be interpreted as measuring the absolute difference between AVQITE and exact ITE in units of the exact thermal expectation value $\braket{\mathcal O}_{\rm ED}$, which is more natural than $\braket{\mathcal O}_{\rm ITE}$ itself as this quantity can be close to zero even when $\braket{\mathcal O}_{\rm ED}$ is finite. (Note that the average value of $\braket{\mathcal O}_{\rm ITE}$ is not the same as the METTS estimate of $\braket{\mathcal O}$ because it samples the METTS uniformly.)
For this metric, we consider as observables the energy density $\epsilon$ [calculated with respect to the Hamiltonian~\eqref{eq:hamiltonian}] and the particle density $n$ [calculated with respect to the number operator~\eqref{eq:number_op}].

\begin{figure*}
     \includegraphics[width=0.9\textwidth]{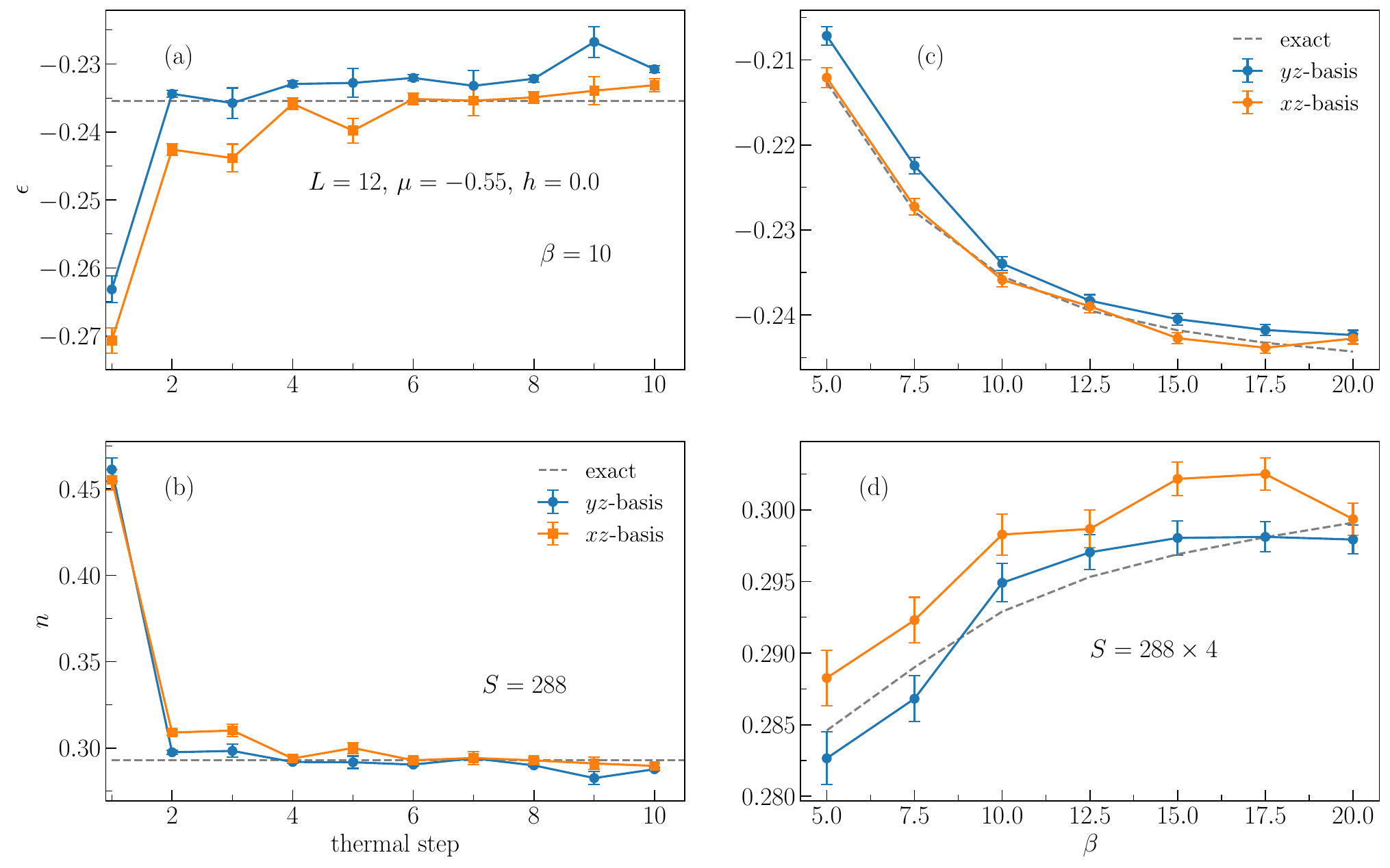}
     \caption{Minimal autocorrelation between thermal steps and accuracy of thermal energy and particle density from AVQMETTS calculations on the model \eqref{eq:hamiltonian} at $L=12$ without a confining field ($h=0$), and with a chemical potential set to $\mu=0.55$, which gives 1/3 particle filling. (a) Thermal energy density $\epsilon$ as a function of thermal steps using the alternating $yz$-basis (blue circles) at $\beta=10$. The results using $xz$-basis (orange squares) are also shown for comparison, together with the exact thermal energy density (dashed line). Each thermal step result represents an average over $S_w = 288$ independent random walks.w
     (b) Same as (a), but for particle density $n$. 
     (c) Thermal energy density as a function of $\beta$ from AVQMETTS calculations using $yz$- and $xz$-bases, in comparison with exact results.
     The AVQMETTS result at each temperature is an average over $S_w\times S_0 = 288\times 4$ samples.
     (d) Same as (c) but for particle density.
     }
     \label{fig:plt_e_and_n}
\end{figure*}

Figure~\ref{fig:plt_acc} shows the distribution of the above figures of merit over the $288$ initial CPSs per product state basis for inverse temperatures $\beta=1$ (top panels) and $\beta=2$ (bottom panels).
While all the infidelities are smaller than $10^{-2}$ as shown in Fig.~\ref{fig:plt_acc}(a,b), the infidelities of the METTSs from AVQITE calculations are smallest when the initial CPSs are $z$-basis states, followed by those in the $y$- and $x$-basis, respectively.
When $\beta$ increases from $1$ to $2$, the infidelities generally increase as well due to error accumulations from taking more imaginary-time steps of similar step size.
Nevertheless, this trend of increasing infidelities with $\beta$ does not necessarily persist to low temperatures, as demonstrated in Ref.~\cite{Getelina23}, which is also corroborated by the accurate AVQMETTS calculations at low temperatures ($\beta=5, 10, 20$) to be reported later.
In Fig.~\ref{fig:plt_acc}(c) we show the distributions of $\tilde\Delta_E$ for AVQITE calculations starting from the three initial CPS bases, all of which roughly peak around $10^{-1.8} \approx 1.6\%$ at $\beta=1$. 
At larger $\beta=2$ as shown in Fig.~\ref{fig:plt_acc}(d), the distribution of $\tilde\Delta_E$ of the three groups are more separated, where the peak position shifts to $10^{-1.75} \approx 1.8\%$ for $z$-basis results, $10^{-1.7} \approx 2.0\%$ for $y$-basis, and $10^{-1.6} \approx 2.5\%$ for $x$-basis. 
The trend correlates well with that of the infidelities. 
Fig.~\ref{fig:plt_acc}(e,f) plots the distributions of $\tilde\Delta_N$ for $\beta=1$ and $2$, respectively.
Here, the $z$-basis results have significantly smaller error as compared to the $x/y$-basis results. 
This can be understood by noting that the initial CPSs in the $z$-basis are eigenstates of the particle number operator $N$, which also commutes with the Hamiltonian $H$. 
Therefore, in principle, ideal ITE under $H$ only propagates these states within a fixed-particle number sector. 
Numerically, about $25\%$ of the 288 AVQITE results in the $z$-basis yield $\tilde\Delta_N=0$ (these values are not included in the histograms). 
The small but finite values of $\tilde \Delta_N$ for the $z$-basis results in Fig.~\ref{fig:plt_acc}(e,f) imply a slight symmetry-breaking in the AVQITE calculations. 
The $\tilde\Delta_N$ distribution of $x$-basis results appears broader than for the $y$-basis results, but with a roughly similar peak position. 
Overall, the particle number results are more accurate than the energy results.

\begin{figure*}
     \includegraphics[width=0.9\textwidth]{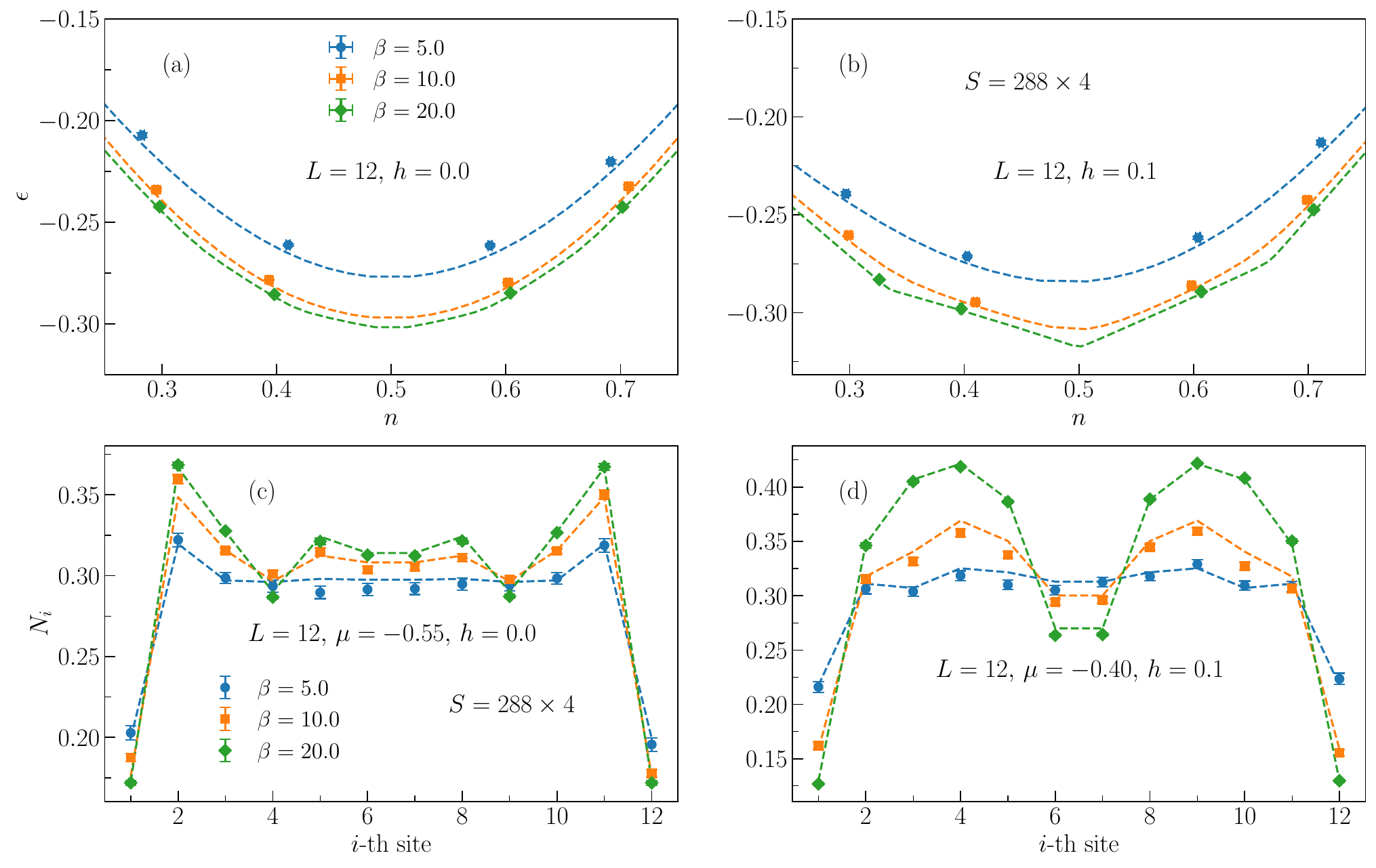}
     \caption{AVQMETTS calculations of energy density and particle number distribution. (a) Energy density $\epsilon$ as a function of particle number density $n$ for a free-fermion system (h=0) of size $L=12$ at inverse temperatures $\beta= 5, 10, 20$. (b) Same as (a) but for a finite confining field $h=0.1$. 
     We choose chemical potential values $\mu \in (-0.6, 0.6)$ to get 4 points (symbols) at each temperature and compare them with exact calculations (dashed curves).
     (c) Particle occupation $n_i$ as a function of site $i$ for $L=12$ and $h=0$ from AVQMETTS calculations (symbols), with comparison to exact results (dashed lines). (d) Same as (c) but for $h=0.1$.
     The chemical potential is set to $\mu=-0.55$ in (c) and $\mu=-0.40$ in (d) to yield $1/3$ particle filling in the ground state. Each AVQMETTS calculation generates $S_w \times S_0 = 288\times 4$ samples for thermal averaging.}
     \label{fig:plt_eq_state_friedel}
\end{figure*}

From the previous discussion one may argue in favor of AVQMETTS simulations using only the $z$ basis for state collapse, as Fig.~\ref{fig:plt_acc} demonstrates that AVQITE simulations for CPSs in the $z$ basis give the best accuracy overall. 
However, it is also important to consider the sampling efficiency for the thermal average of an observable $\mathcal{O}$, which is tied to the spread of the distribution of $\langle \mathcal{O} \rangle_i$ over METTSs with respect to the exact thermal average $\langle \mathcal{O} \rangle_\textrm{ED}$. 
High sampling efficiency is expected if the distribution is narrowly peaked around $\langle \mathcal{O} \rangle_\textrm{ED}$. 
Note that, as mentioned above, the average of $\langle \mathcal{O} \rangle_i$ over METTS obtained from $\mathcal M$ uniformly sampled CPSs need not converge to $\langle \mathcal{O} \rangle_\textrm{ED}$.
We therefore define the following measure of relative error with respect to the exact thermal average:
\begin{equation}
    \mathcal{D}_\mathcal{O} = \frac{1}{\mathcal{M}}\sum_{i=1}^\mathcal{M} \left| \left(\mathcal{O}_i-\langle\mathcal{O}\rangle_\textrm{ED}\right)/\langle\mathcal{O}\rangle_\textrm{ED }\right| \,,\label{eq:error_ed}
\end{equation}
which measures the spread of the distribution of $\langle \mathcal{O}\rangle_i$ relative to $\langle \mathcal{O} \rangle_\textrm{ED}$.

Table~\ref{tab:analysis_ed} shows the spread of the energies $\langle H \rangle_i$ and particle numbers $\langle N \rangle_i$ characterized by $\mathcal{D}_E$ and $\mathcal{D}_N$ from both AVQITE (bold numbers) and exact-ITE calculations. 
We used the same $\mathcal{M}=288$ states as in Fig.~\ref{fig:plt_acc} to generate this table. 
The values obtained from AVQITE calculations are very close to those from exact ITE, which further corroborates the accuracy of AVQITE. 
Across two representative temperatures ($\beta=1, 2$), one can see that calculations using the $y$ basis consistently produce the smallest $\mathcal{D}_E$ and $\mathcal{D}_N$, while results using $z$($x$)-basis states are smaller for $\mathcal{D}_E$ ($\mathcal{D}_N$). 

\begin{table}[h]
    \centering
    \begin{tabular}{lcccccl}
     \toprule
     & \multicolumn{3}{c}{$\mathcal{D}_E (\%)$}& \multicolumn{3}{c}{$\mathcal{D}_N (\%)$}
\\\cmidrule(lr){2-4}\cmidrule(lr){5-7}
     & $X$  & $Y$ & $Z$    & $X$  & $Y$ & $Z$\\\midrule
     $\beta=1.0$ & \begin{tabular}{@{}c@{}}$\mathbf{4.06}$ \\ 3.99\end{tabular} & \begin{tabular}{@{}c@{}}$\mathbf{0.40}$ \\ 0.38\end{tabular} & \begin{tabular}{@{}c@{}}$\mathbf{1.53}$ \\ 1.53\end{tabular} & \begin{tabular}{@{}c@{}}$\mathbf{0.29}$ \\ 0.29\end{tabular} & \begin{tabular}{@{}c@{}}$\mathbf{0.09}$ \\ 0.09\end{tabular} & \begin{tabular}{@{}c@{}}$\mathbf{2.80}$ \\ 2.80\end{tabular}
     \\
     $\beta=2.0$ & \begin{tabular}{@{}c@{}}$\mathbf{2.25}$ \\ 2.12\end{tabular} & \begin{tabular}{@{}c@{}}$\mathbf{0.47}$ \\ 0.42\end{tabular} & \begin{tabular}{@{}c@{}}$\mathbf{1.40}$ \\ 1.42\end{tabular} & \begin{tabular}{@{}c@{}}$\mathbf{0.96}$ \\ 0.92\end{tabular} & \begin{tabular}{@{}c@{}}$\mathbf{0.17}$ \\ 0.17\end{tabular} & \begin{tabular}{@{}c@{}}$\mathbf{4.17}$ \\ 4.17\end{tabular}
     \\
     \bottomrule
    \end{tabular}
    \caption{The spread of energy and particle number distributions for METTSs with respect to exact thermal averages, as measured by $\mathcal{D}_E$ and $\mathcal{D}_N$~\eqref{eq:error_ed}, at two different temperatures $\beta=1,2$. We choose random samples of $\mathcal{M} = 288$ CPSs in each of the three ($x/y/z$)-basis representations, which are evolved up to $\beta/2$ using both AVQITE and exact-ITE for the analysis.}
    \label{tab:analysis_ed}
\end{table}

Even though the AVQITE simulations for the CPSs in the $y$ basis yields the narrower energy and particle number distributions relative to the exact thermal averages (which implies better sampling efficiency for AVQMETTS), in practice, alternating the measurement basis for METTS collapse is found to reduce autocorrelation effects between thermal steps~\cite{stoudenmire2010,Getelina23}. 
Given the better AVQITE simulation accuracy for CPSs in the $z$-basis as demonstrated in Table~\ref{tab:analysis_ed} and the associated smaller $\mathcal{D}_E$, we choose the $yz$-basis for AVQMETTS, where calculations alternate between $y$- and $z$-basis measurements at odd and even thermal steps, respectively. 
To demonstrate the effectiveness of the $yz$-basis for AVQMETTS simulations, we plot in Fig.~\ref{fig:plt_e_and_n}(a,b) the energy and particle number densities as functions of the number of thermal steps at $\beta=10$. 
Minimal autocorrelation effects are observed, as the thermal averages already converge to the exact thermal expectation values (dashed lines) at the second thermal step subject to fluctuations tied to the sample size $S=288$.
This suggests that it suffices to skip the first thermal step for each Markovian random walk in AVQMETTS to get a METTS ensemble to estimate thermal averages. 
We follow this procedure for the following calculations.
For reference, we also show the results with the $xz$-basis as adopted in Refs.~\cite{stoudenmire2010,Getelina23}, where the calculations use an alternating $x$- and $z$-basis for collapse. 
One can see a faster convergence for $yz$-basis calculations compared with $xz$ results, especially for the energy density.
In Fig.~\ref{fig:plt_e_and_n}(c,d) we show the AVQMETTS estimates of thermal energy and particle density as functions of $\beta$ with $S=288\times 4$ samples. 
The data from both collapse bases ($yz$ and $xz$) agree well with the exact results. The maximum relative error for the energy density is about $2.6\%$ for the $yz$-basis and $2.0\%$ for the $xz$-basis, whereas the relative error for the particle density is $0.8\%$ for the former and $2.5\%$ for the latter.


\subsection{AVQMETTS simulations of energy and particle densities}

With the important simulation parameters for AVQMETTS optimized through the analyses above, we proceed to benchmark calculations of the equation of state and Friedel oscillations of the spatial particle density distribution. 

In Fig.~\ref{fig:plt_eq_state_friedel}(a) we plot the energy density as a function of the particle number density for the model \eqref{eq:hamiltonian} at $L=12$ and without confining field ($h=0$). 
The AVQMETTS results (symbols) agree well with the exact diagonalization results (dashed curves) at three representative temperatures $\beta=5, 10, 20$. 
The deviations become slightly bigger when temperature increases ($\beta$ reduces), which can be attributed to the fact that we use a fixed ensemble size of $S_w\times S_0 = 288\times 4$.
The maximal relative error is estimated to be $\Delta_\epsilon = 3\%$ at $\beta=5$ by aligning $n$ with the exact result. 
As discussed in Sec.~\ref{sec: eos}, the finite confining field breaks the particle-hole symmetry of the model, which might make the AVQMETTS simulations more challenging. Nevertheless, as we show in Fig.~\ref{fig:plt_eq_state_friedel}(b), the accuracy of energy and particle density remains equally good with finite confining field $h=0.1$. 
In these calculations, we selected four values of the chemical potential in $\mu \in [-0.6, 0.6]$ to obtain particle density points in the vicinity of half filling ($n=0.5$).

We further present the spatial distribution of the particle density $\langle n_i \rangle$ of the model without confinement in Fig.~\ref{fig:plt_eq_state_friedel}(c), which shows Friedel oscillations with a number of peaks equal to the particle occupancy $\langle N \rangle$, as discussed in Sec.~\ref{sec: FO}. 
Here one can verify that the AVQMETTS results (symbols) also agree well with the exact results (dashed lines), where the four peaks are well resolved. 
As above, the deviation becomes slightly bigger with increasing temperature due to our use of a fixed ensemble size. 
At $\beta=5$ where the oscillations are almost washed out, the maximal relative error is about $\Delta_{n_i} = 4.2\%$. 
Here we set $\mu=-0.55$ such that the system has $4$ particles in the ground state, consistent with calculations in Sec.~\ref{sec: FO}. 
We also find a consistent accuracy for the particle density distribution with finite confining field ($h=0.1$), where the number of peaks reduces to two due to particle pairing, as shown in Fig.~\ref{fig:plt_eq_state_friedel}(d).

\begin{figure*}[t]
     \includegraphics[width=\textwidth]{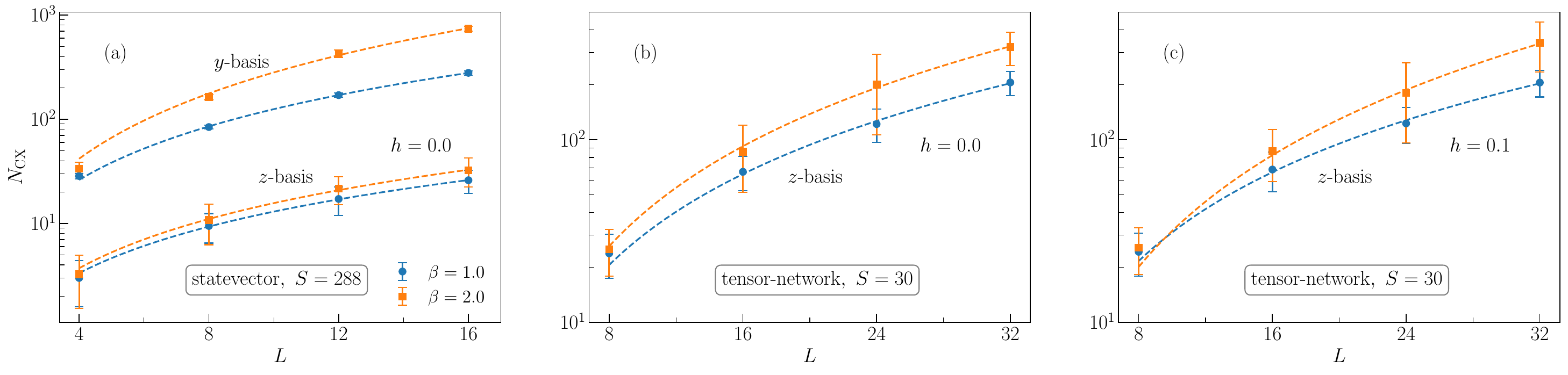}
     \caption{System-size dependence of the AVQMETTS ansatz circuit complexity as measured by the number of CNOT gates $N_\textrm{CX}$. 
     (a) Growth of $N_\textrm{CX}$ as a function of the system size $L$ for the deconfined free-fermion model with $h=0$. The results are shown for the two alternating $y$- and $z$-measurement basis for METTS collapse at temperatures $\beta=1, 2$. Each data point corresponds to an average over 288 random CPSs obtained using statevector simulations, while the error bars indicate the standard deviation of the corresponding distribution.
     (b) Same as (a) but the averaging is performed over 30 random CPSs obtained using tensor network simulations.
     We checked that increasing the sample size to 60 random CPSs does not meaningfully alter the results.
     (c) Same as (b) but with the confining field $h=0.1$.
     In (a)-(c) we select a chemical potential $\mu$ for each system size to maintain a total filling $\langle N \rangle=L/4$ in the ground state.
     Dashed lines are best fits of the data with the function $f(L)=aL^b$.}
     \label{fig:plt_ncx}
\end{figure*}

\subsection{System-size scaling of AVQMETTS circuit complexity}

Even though the AVQMETTS calculations above are performed on classical statevector simulator, we have access to the circuit representation and can therefore estimate the circuit complexity characterized by the number of CNOT gates $N_{\textrm{CX}}$ used in the variational ansatz circuits at the end of the ITE process. 
For simplicity, we assume here a QPU with all-to-all qubit connectivity, which applies to trapped-ion devices. 
Thereby, for a given AVQITE ansatz with $N_\theta$ unitaries each of which has an angle $\theta_j$ associated with a Pauli string $G_j$ as a generator, we can estimate $N_{\textrm{CX}}$ as:
\begin{equation}
    N_\textrm{CX} = \sum_{j=1}^{N_\theta} 2 \left(W_j - 1\right),\label{eq:ncx}
\end{equation}
where $W_j$ is the weight (number of nonidentity Pauli operators) of the Pauli string $G_j$. 
Notice that Eq.~\eqref{eq:ncx} sets an upper bound for the total number of CNOT gates: it does not account for possible gate cancellations in the translation of the ansatz to a quantum circuit.

In Fig.~\ref{fig:plt_ncx}(a) we show the statistics of $N_{\textrm{CX}}$ (symbols) for AVQITE circuits with initial CPSs in the $y$ and $z$ bases as a function of system size $L$. 
Here the confining field is set to $h=0$, and the chemical potential $\mu$ is adjusted such that the total fermion filling $\langle N \rangle = L/4$. 
We apply AVQITE to propagate the CPSs up to temperatures $\beta=1, 2$. 
The dashed lines show that the data are fitted well by a simple power-law function $f(L)=aL^b$, with exponent $b$ ranging from $1.6$ to $2.1$ for the different measurement basis and temperature values considered here. 
One can clearly see a bimodal distribution of AVQITE calculations: for calculations starting from $z$-basis CPSs, $N_{\textrm{CX}}$ is in the range of $[4, 30]$, while for ones starting from $y$-basis CPSs it is in the range $[20, 800]$, demonstrating over an order of magnitude increase at fixed $L$. 
Furthermore, increasing $\beta$ at fixed $L$ leads to only a slight increase in $N_{\textrm{CX}}$ for $z$-basis results, in contrast to the substantially larger increase observed for the $y$-basis results. 
Hence, even though we showed in Table~\ref{tab:analysis_ed} that the $y$-basis samples are more concentrated around the thermal average, the implied greater sampling efficiency can be compromised by the increased quantum resource overhead in AVQMETTS calculations. 
This again suggests that AVQMETTS using the alternating $yz$-basis for state collapse may be an optimal choice.

In Fig.~\ref{fig:plt_ncx}(b) and (c) we extend the results for $N_{\textrm{CX}}$ to larger system sizes (up to $L=32$) using a tensor-network (TN) simulator of the AVQITE algorithm~\cite{Khindanov2025inpreparation} based on the \texttt{Quimb}~\cite{gray2018quimb} and \texttt{cotengra} packages~\cite{Gray2021hyperoptimized}.
Unlike the statevector simulations discussed above, where we sampled over $288$ initial CPSs, here we instead sample over $30$ initial CPSs.
Nevertheless, we have checked that doubling the number of samples to $60$ does not meaningfully alter the results.
In addition, due to the higher circuit depths for the $y$-basis samples, we perform TN simulations for the initial CPSs in the $z$-basis only.
However, there is no obstruction to performing the $y$-basis calculations, increasing the number of samples, and considering system sizes larger than $L=32$ given sufficient HPC resources.

Fig.~\ref{fig:plt_ncx}(b) demonstrates that for the larger system sizes, the means of the $N_{\textrm{CX}}$ distributions still follow the power-law dependence $f(L)=aL^b$, with the exponent $b$ ranging between $1.6$ and $2.1$.
Additionally, Fig.~\ref{fig:plt_ncx}(c) shows that the power-law dependence of $N_{\textrm{CX}}$ persists as we change the confining field from $h=0$ to $h=0.1$. 
Note that the mismatch in $N_{\textrm{CX}}$ between Fig.~\ref{fig:plt_ncx}(a) and (b) at $L=8$ and $L=16$ comes from the fact that in (a) the implementation of AVQITE prioritizes the selection of low-weight generators if their scores are close, whereas in (b) this is not the case and the generators with close scores are sorted in alphabetical order. We, however, do not expect such fine tunings of the algorithm to affect the scaling behavior of $N_{\textrm{CX}}$, and indeed panels (a) and (b) yield consistent estimates of this scaling. 

These results showcase that TN-based simulations can serve as a promising avenue for benchmarking and analyzing the system-size scaling complexity of quantum algorithms for system sizes beyond the reach of statevector simulators.
This is especially relevant for heuristic quantum algorithms (such as variational ones), for which scaling behavior and use-case-appropriate bounds on other resource requirements generally cannot be obtained through analytical arguments~\cite{Zimboras2025}.

In addition, the TN simulator can be leveraged by combining it with a QPU in a hybrid quantum-classical workflow for the AVQITE algorithm.
In such a workflow, the TN approach would be used to classically simulate quantum circuits that are amenable to such simulations (e.g., circuits with shallower ans\"{a}tze that comprise a block of the quantum geometric tensor corresponding to the early stages of the algorithm), while a QPU would only be utilized for circuits whose complexity is inaccessible to state of the art TN techniques.
Crucially, for any given quantum circuit, the computational and memory requirements of its TN contraction (i.e. the contraction cost and width) can be estimated \textit{prior} to evaluating the contraction, from which the feasibility of such classical evaluation can be assessed.
The described hybrid TN-QPU workflow can considerably reduce the number of QPU circuit evaluations needed to scale AVQITE to the beyond-classical regime~\cite{Zhang2025}.



\subsection{Alternative Methods}

Having presented a detailed benchmark of the AVQMETTS approach to simulating finite-temperature and -density properties of the 1+1 dimensional $\mathbb Z_2$ lattice gauge theory, we briefly survey some alternative approaches to the problem.
In general, simulation of finite temperature physics on a quantum computer is more involved than zero-temperature as quantum computers do not naturally operate on mixed states.

The METTS approach explored in this work can be used in combination with any algorithm implementing the imaginary time evolution primitive on a quantum computer.
Besides the variational approach adopted here, several other implementations of the ITE primitive have been proposed.
Ref.~\cite{Motta19} proposes to perform ITE with the help of state tomography on a neighborhood whose size scales with the correlation length.
ITE can also be implemented using a register of ancilla qubits as well as mid-circuit measurements or resets, as in Refs.~\cite{Shtanko23,Mao23}.
More broadly, ITE can be implemented using generic primitives such as quantum signal processing~\cite{Low16,Low17,Low19,Martyn21,Coopmans23} or block encoding~\cite{Poulin09,Chowdhury17,Gilyen19,vanApeldoorn20,Chowdhury21}.

There are also alternative approaches to estimating thermal averages on quantum computers. 
Some of these assume access to ITE as a primitive.
The ``typicality" based methods can compute finite-temperature averages by evolving a fully random or pseudo-random state~\cite{Coopmans23,Davoudi23} of the entire system with the preparation of this pseudo-random state being efficient in principle. 
Another approach gives access to a purification of the full thermal density matrix by performing imaginary time evolution on half of a collection of Bell pairs~\cite{Yuan19}.
Finally, one can also consider fully variational methods for preparing mixed states by parameterizing $\rho(\boldsymbol{\theta},\boldsymbol{\phi}) = \sum_k p_k(\boldsymbol{\theta}) \ket{\psi_k(\boldsymbol{\phi})}\bra{\psi_k(\boldsymbol{\phi})}$ and minimizing the free energy~\cite{Verdon19}. 
Due to the difficulty of measuring the entropic term, other variants of this scheme have been proposed using, for example, mid-circuit measurements~\cite{Selisko23}. 
Microcanonical averages may also be accessible by generating variational samples of the microcanonical ensemble \cite{Pollock23}.

As with AVQMETTS, the scalability of these various approaches in practice for near-term hardware is not yet established. 
It will be necessary to continue investigating means to reduce obstacles such as measurement, qubit, and gate overheads for all of these approaches. 
Furthermore, these methods are all impacted in different ways by gate errors, measurement errors, and shot noise.
As hardware and algorithms continue to evolve, careful benchmarks like the ones performed in this paper are necessary to gauge the feasibility of finite-temperature quantum simulation on quantum computers.

\section{Conclusion}
\label{sec:outlook}

In this paper we investigated the utility of METTS approaches to the simulation of gauge theories at finite temperature and density.
The METTS approach is useful both in classical simulations of quantum matter via tensor network approaches and is a promising candidate for implementing finite-temperature simulations on quantum computers via, e.g., AVQMETTS.
We focused on the simplest lattice model of a gauge theory coupled to matter, the one-dimensional $\mathbb Z_2$ gauge theory, and demonstrated the applicability of METTS techniques to study the finite-temperature and -density equation of state as well as signatures of fermion confinement.
We investigated systematic issues common to both the classical and quantum implementations of METTS, particularly the impact of CPS collapse basis on the sampling complexity.
We also benchmarked the impact of sampling basis on the AVQITE quantum ITE subroutine in terms of both simulation accuracy and quantum resource cost.
Our study demonstrates that METTS approaches are capable of accurately simulating the confined and deconfined regimes of the model in a large range of temperatures and system sizes.
We performed classical METTS simulations of 1D chains at system sizes exceeding those accessible to purification approaches~\cite{Kebrič24}, as well as of smaller systems more likely to be the subject of near-term quantum computing studies.
We find that an optimal choice of sampling basis can speed up convergence of the METTS sampling, but also demonstrate a cautionary observation that the best sampling basis may lead to unfavorable circuit complexity scaling in a quantum computing setting.
We suggest an alternating collapse basis approach to exploit the advantages of different bases that we expect will prove useful in future studies.

Although we have focused in this paper on the simplest 1+1-dimensional $\mathbb Z_2$ lattice gauge theory, we note that many of the techniques adopted here can potentially be applied to higher-dimensional systems and for other gauge groups.
For example, a 2+1 dimensional version of the $\mathbb Z_2$ lattice gauge theory studied in this work has recently been shown to map to a two-dimensional qubit-only model~\cite{Borla22}.
Spin-1/2 formulations of $U(1)$ lattice gauge theory in 1+1 and 2+1 dimensions~\cite{Surace19,Celi20} are amenable to digital quantum simulation with qubits, and some initial progress has been made in understanding the confining dynamics in the 1+1 dimensional case~\cite{Chen21}.
The 2+1 dimensional models can be studied using METTS on narrow cylinders, as has recently been done for some models of correlated electron systems~\cite{Wietek21a,Wietek21b}.
There are also prospects for simulating non-Abelian gauge theories in 1+1~\cite{Silvi19,Rigobello23} and 2+1 dimensions~\cite{Cataldi24} using tensor networks, and for quantum simulations of such models in, e.g., trapped-ion quantum computers~\cite{Calajo24}.
More generally, encodings into qubit- or qudit-only models have been proposed for U($N$) and SU($N$) gauge theories with fermionic matter in 2+1 dimensions~\cite{Zohar18}.
For the simplest of these models (e.g.~the Abelian ones), quantum simulation of finite-temperature properties using the methods explored in this work may soon be feasible.
For the more complicated models with tensor-network formulations, finite-temperature classical simulations using METTS are a natural next step.

The methods adopted in this work to calculate the equation of state may offer insight into two topics of current interest in the study of neutron matter, namely neutron star mass-radius relations~\cite{Watts:2016uzu, Ozel:2016oaf, Demorest:2010bx, Antoniadis:2013pzd} and multi-messenger signals of neutron star mergers~\cite{Radice:2017lry,PhysRevLett.108.011101,Takami:2014zpa, Radice:2016rys, Chatziioannou:2017ixj}. 
Both require a knowledge of the equation of state of nuclear matter at finite density and temperature. 
In the case of the former, neutron star observational data which favor a soft equation of state at low density and a stiff one at high density are in tension with several theoretical models which predict a rather soft equation of state at high density \cite{Bedaque:2014sqa}. 
In fact this soft-stiff equation of state data from neutron stars suggests that the speed of sound is non-monotonic as a function of density.
Since first-principles QCD calculations are currently out of reach, it is worthwhile exploring whether toy models in lower dimensional theories like the one studied in this work can exhibit some of these features expected in the QCD equation of state. 
Similarly, nuclear/QCD equations of state at relatively higher temperature are used in numerical relativity simulations to predict gravitational wave signals from neutron star merger events. 
Computing finite temperature equations of state for toy models is a step towards eventually doing the same for QCD.
By investigating the systematics of (AVQ)METTS calculations, our paper lays the groundwork for both classical and quantum computers to tackle these questions in the future.

\section*{Data availability}
Data used in all plots in this manuscript are available at \url{https://gitlab.com/gqce/avqmetts_z2}.

\section*{Code availability}
Code used to produce all plots in this manuscript is available at \url{https://gitlab.com/gqce/avqmetts_z2}.

\acknowledgments
The authors acknowledge valuable discussions with Matja\v{z} Kebri\v{c} and Peter P. Orth.
This work was supported by the U.S. Department of Energy (DOE), Office of Science, Basic Energy Sciences, Materials Science and Engineering Division, including the grant of computer time at the National Energy Research Scientific Computing Center (NERSC) in Berkeley, California. The research was performed at the Ames National Laboratory, which is operated for the U.S. DOE by Iowa State University under Contract No. DE-AC02-07CH11358. S.S. acknowledges support from the U.S. Department of Energy, Nuclear Physics Quantum Horizons program through the Early Career Award DE-SC0021892. 

\bibliography{Refs}

\begin{thebibliography}{95}%
\makeatletter
\providecommand \@ifxundefined [1]{%
 \@ifx{#1\undefined}
}%
\providecommand \@ifnum [1]{%
 \ifnum #1\expandafter \@firstoftwo
 \else \expandafter \@secondoftwo
 \fi
}%
\providecommand \@ifx [1]{%
 \ifx #1\expandafter \@firstoftwo
 \else \expandafter \@secondoftwo
 \fi
}%
\providecommand \natexlab [1]{#1}%
\providecommand \enquote  [1]{``#1''}%
\providecommand \bibnamefont  [1]{#1}%
\providecommand \bibfnamefont [1]{#1}%
\providecommand \citenamefont [1]{#1}%
\providecommand \href@noop [0]{\@secondoftwo}%
\providecommand \href [0]{\begingroup \@sanitize@url \@href}%
\providecommand \@href[1]{\@@startlink{#1}\@@href}%
\providecommand \@@href[1]{\endgroup#1\@@endlink}%
\providecommand \@sanitize@url [0]{\catcode `\\12\catcode `\$12\catcode `\&12\catcode `\#12\catcode `\^12\catcode `\_12\catcode `\%12\relax}%
\providecommand \@@startlink[1]{}%
\providecommand \@@endlink[0]{}%
\providecommand \url  [0]{\begingroup\@sanitize@url \@url }%
\providecommand \@url [1]{\endgroup\@href {#1}{\urlprefix }}%
\providecommand \urlprefix  [0]{URL }%
\providecommand \Eprint [0]{\href }%
\providecommand \doibase [0]{https://doi.org/}%
\providecommand \selectlanguage [0]{\@gobble}%
\providecommand \bibinfo  [0]{\@secondoftwo}%
\providecommand \bibfield  [0]{\@secondoftwo}%
\providecommand \translation [1]{[#1]}%
\providecommand \BibitemOpen [0]{}%
\providecommand \bibitemStop [0]{}%
\providecommand \bibitemNoStop [0]{.\EOS\space}%
\providecommand \EOS [0]{\spacefactor3000\relax}%
\providecommand \BibitemShut  [1]{\csname bibitem#1\endcsname}%
\let\auto@bib@innerbib\@empty
\bibitem [{\citenamefont {Stephanov}(2006)}]{Stephanov:2006zvm}%
  \BibitemOpen
  \bibfield  {author} {\bibinfo {author} {\bibfnamefont {M.~A.}\ \bibnamefont {Stephanov}},\ }\bibfield  {title} {\bibinfo {title} {{QCD phase diagram: An Overview}},\ }\href {https://doi.org/10.22323/1.032.0024} {\bibfield  {journal} {\bibinfo  {journal} {PoS}\ }\textbf {\bibinfo {volume} {LAT2006}},\ \bibinfo {pages} {024} (\bibinfo {year} {2006})},\ \Eprint {https://arxiv.org/abs/hep-lat/0701002} {arXiv:hep-lat/0701002} \BibitemShut {NoStop}%
\bibitem [{\citenamefont {Sorensen}\ \emph {et~al.}(2024)\citenamefont {Sorensen} \emph {et~al.}}]{Sorensen:2023zkk}%
  \BibitemOpen
  \bibfield  {author} {\bibinfo {author} {\bibfnamefont {A.}~\bibnamefont {Sorensen}} \emph {et~al.},\ }\bibfield  {title} {\bibinfo {title} {{Dense nuclear matter equation of state from heavy-ion collisions}},\ }\href {https://doi.org/10.1016/j.ppnp.2023.104080} {\bibfield  {journal} {\bibinfo  {journal} {Prog. Part. Nucl. Phys.}\ }\textbf {\bibinfo {volume} {134}},\ \bibinfo {pages} {104080} (\bibinfo {year} {2024})},\ \Eprint {https://arxiv.org/abs/2301.13253} {arXiv:2301.13253 [nucl-th]} \BibitemShut {NoStop}%
\bibitem [{\citenamefont {Rajagopal}(1999)}]{Rajagopal:1999cp}%
  \BibitemOpen
  \bibfield  {author} {\bibinfo {author} {\bibfnamefont {K.}~\bibnamefont {Rajagopal}},\ }\bibfield  {title} {\bibinfo {title} {{Mapping the QCD phase diagram}},\ }\href {https://doi.org/10.1016/S0375-9474(99)85017-9} {\bibfield  {journal} {\bibinfo  {journal} {Nucl. Phys. A}\ }\textbf {\bibinfo {volume} {661}},\ \bibinfo {pages} {150} (\bibinfo {year} {1999})},\ \Eprint {https://arxiv.org/abs/hep-ph/9908360} {arXiv:hep-ph/9908360} \BibitemShut {NoStop}%
\bibitem [{\citenamefont {Wen}(2004)}]{wen2004quantum}%
  \BibitemOpen
  \bibfield  {author} {\bibinfo {author} {\bibfnamefont {X.-G.}\ \bibnamefont {Wen}},\ }\href@noop {} {\emph {\bibinfo {title} {Quantum field theory of many-body systems: From the origin of sound to an origin of light and electrons}}}\ (\bibinfo  {publisher} {Oxford university press},\ \bibinfo {year} {2004})\BibitemShut {NoStop}%
\bibitem [{\citenamefont {Senthil}\ \emph {et~al.}(2004)\citenamefont {Senthil}, \citenamefont {Vishwanath}, \citenamefont {Balents}, \citenamefont {Sachdev},\ and\ \citenamefont {Fisher}}]{Senthil2004}%
  \BibitemOpen
  \bibfield  {author} {\bibinfo {author} {\bibfnamefont {T.}~\bibnamefont {Senthil}}, \bibinfo {author} {\bibfnamefont {A.}~\bibnamefont {Vishwanath}}, \bibinfo {author} {\bibfnamefont {L.}~\bibnamefont {Balents}}, \bibinfo {author} {\bibfnamefont {S.}~\bibnamefont {Sachdev}},\ and\ \bibinfo {author} {\bibfnamefont {M.~P.~A.}\ \bibnamefont {Fisher}},\ }\bibfield  {title} {\bibinfo {title} {Deconfined quantum critical points},\ }\href {https://doi.org/10.1126/science.1091806} {\bibfield  {journal} {\bibinfo  {journal} {Science}\ }\textbf {\bibinfo {volume} {303}},\ \bibinfo {pages} {1490–1494} (\bibinfo {year} {2004})}\BibitemShut {NoStop}%
\bibitem [{\citenamefont {Assaad}\ and\ \citenamefont {Grover}(2016)}]{Grover16}%
  \BibitemOpen
  \bibfield  {author} {\bibinfo {author} {\bibfnamefont {F.~F.}\ \bibnamefont {Assaad}}\ and\ \bibinfo {author} {\bibfnamefont {T.}~\bibnamefont {Grover}},\ }\bibfield  {title} {\bibinfo {title} {Simple fermionic model of deconfined phases and phase transitions},\ }\href {https://doi.org/10.1103/PhysRevX.6.041049} {\bibfield  {journal} {\bibinfo  {journal} {Phys. Rev. X}\ }\textbf {\bibinfo {volume} {6}},\ \bibinfo {pages} {041049} (\bibinfo {year} {2016})}\BibitemShut {NoStop}%
\bibitem [{\citenamefont {Wegner}(1971)}]{Wegner71}%
  \BibitemOpen
  \bibfield  {author} {\bibinfo {author} {\bibfnamefont {F.~J.}\ \bibnamefont {Wegner}},\ }\bibfield  {title} {\bibinfo {title} {Duality in generalized ising models and phase transitions without local order parameters},\ }\href {https://doi.org/10.1063/1.1665530} {\bibfield  {journal} {\bibinfo  {journal} {Journal of Mathematical Physics}\ }\textbf {\bibinfo {volume} {12}},\ \bibinfo {pages} {2259–2272} (\bibinfo {year} {1971})}\BibitemShut {NoStop}%
\bibitem [{\citenamefont {Kogut}(1979)}]{Kogut79}%
  \BibitemOpen
  \bibfield  {author} {\bibinfo {author} {\bibfnamefont {J.~B.}\ \bibnamefont {Kogut}},\ }\bibfield  {title} {\bibinfo {title} {An introduction to lattice gauge theory and spin systems},\ }\href {https://doi.org/10.1103/RevModPhys.51.659} {\bibfield  {journal} {\bibinfo  {journal} {Rev. Mod. Phys.}\ }\textbf {\bibinfo {volume} {51}},\ \bibinfo {pages} {659} (\bibinfo {year} {1979})}\BibitemShut {NoStop}%
\bibitem [{\citenamefont {Senthil}\ and\ \citenamefont {Fisher}(2000)}]{Senthil00}%
  \BibitemOpen
  \bibfield  {author} {\bibinfo {author} {\bibfnamefont {T.}~\bibnamefont {Senthil}}\ and\ \bibinfo {author} {\bibfnamefont {M.~P.~A.}\ \bibnamefont {Fisher}},\ }\bibfield  {title} {\bibinfo {title} {${Z}_{2}$ gauge theory of electron fractionalization in strongly correlated systems},\ }\href {https://doi.org/10.1103/PhysRevB.62.7850} {\bibfield  {journal} {\bibinfo  {journal} {Phys. Rev. B}\ }\textbf {\bibinfo {volume} {62}},\ \bibinfo {pages} {7850} (\bibinfo {year} {2000})}\BibitemShut {NoStop}%
\bibitem [{\citenamefont {Sedgewick}\ \emph {et~al.}(2002)\citenamefont {Sedgewick}, \citenamefont {Scalapino},\ and\ \citenamefont {Sugar}}]{Sedgewick02}%
  \BibitemOpen
  \bibfield  {author} {\bibinfo {author} {\bibfnamefont {R.~D.}\ \bibnamefont {Sedgewick}}, \bibinfo {author} {\bibfnamefont {D.~J.}\ \bibnamefont {Scalapino}},\ and\ \bibinfo {author} {\bibfnamefont {R.~L.}\ \bibnamefont {Sugar}},\ }\bibfield  {title} {\bibinfo {title} {Fractionalized phase in an $\mathrm{XY}--{Z}_{2}$ gauge model},\ }\href {https://doi.org/10.1103/PhysRevB.65.054508} {\bibfield  {journal} {\bibinfo  {journal} {Phys. Rev. B}\ }\textbf {\bibinfo {volume} {65}},\ \bibinfo {pages} {054508} (\bibinfo {year} {2002})}\BibitemShut {NoStop}%
\bibitem [{\citenamefont {Lee}(2007)}]{Lee07}%
  \BibitemOpen
  \bibfield  {author} {\bibinfo {author} {\bibfnamefont {P.~A.}\ \bibnamefont {Lee}},\ }\bibfield  {title} {\bibinfo {title} {From high temperature superconductivity to quantum spin liquid: progress in strong correlation physics},\ }\href {https://doi.org/10.1088/0034-4885/71/1/012501} {\bibfield  {journal} {\bibinfo  {journal} {Reports on Progress in Physics}\ }\textbf {\bibinfo {volume} {71}},\ \bibinfo {pages} {012501} (\bibinfo {year} {2007})}\BibitemShut {NoStop}%
\bibitem [{\citenamefont {Sachdev}\ and\ \citenamefont {Chowdhury}(2016)}]{Sachdev16}%
  \BibitemOpen
  \bibfield  {author} {\bibinfo {author} {\bibfnamefont {S.}~\bibnamefont {Sachdev}}\ and\ \bibinfo {author} {\bibfnamefont {D.}~\bibnamefont {Chowdhury}},\ }\bibfield  {title} {\bibinfo {title} {The novel metallic states of the cuprates: Topological fermi liquids and strange metals},\ }\href {https://doi.org/10.1093/ptep/ptw110} {\bibfield  {journal} {\bibinfo  {journal} {Progress of Theoretical and Experimental Physics}\ }\textbf {\bibinfo {volume} {2016}},\ \bibinfo {pages} {12C102} (\bibinfo {year} {2016})}\BibitemShut {NoStop}%
\bibitem [{\citenamefont {Wilson}(1974)}]{Wilson74}%
  \BibitemOpen
  \bibfield  {author} {\bibinfo {author} {\bibfnamefont {K.~G.}\ \bibnamefont {Wilson}},\ }\bibfield  {title} {\bibinfo {title} {Confinement of quarks},\ }\href {https://doi.org/10.1103/PhysRevD.10.2445} {\bibfield  {journal} {\bibinfo  {journal} {Phys. Rev. D}\ }\textbf {\bibinfo {volume} {10}},\ \bibinfo {pages} {2445} (\bibinfo {year} {1974})}\BibitemShut {NoStop}%
\bibitem [{\citenamefont {Ceperley}\ \emph {et~al.}(1977)\citenamefont {Ceperley}, \citenamefont {Chester},\ and\ \citenamefont {Kalos}}]{Ceperley77}%
  \BibitemOpen
  \bibfield  {author} {\bibinfo {author} {\bibfnamefont {D.}~\bibnamefont {Ceperley}}, \bibinfo {author} {\bibfnamefont {G.~V.}\ \bibnamefont {Chester}},\ and\ \bibinfo {author} {\bibfnamefont {M.~H.}\ \bibnamefont {Kalos}},\ }\bibfield  {title} {\bibinfo {title} {Monte carlo simulation of a many-fermion study},\ }\href {https://doi.org/10.1103/PhysRevB.16.3081} {\bibfield  {journal} {\bibinfo  {journal} {Phys. Rev. B}\ }\textbf {\bibinfo {volume} {16}},\ \bibinfo {pages} {3081} (\bibinfo {year} {1977})}\BibitemShut {NoStop}%
\bibitem [{\citenamefont {Blankenbecler}\ \emph {et~al.}(1981)\citenamefont {Blankenbecler}, \citenamefont {Scalapino},\ and\ \citenamefont {Sugar}}]{Blankenbecler81}%
  \BibitemOpen
  \bibfield  {author} {\bibinfo {author} {\bibfnamefont {R.}~\bibnamefont {Blankenbecler}}, \bibinfo {author} {\bibfnamefont {D.~J.}\ \bibnamefont {Scalapino}},\ and\ \bibinfo {author} {\bibfnamefont {R.~L.}\ \bibnamefont {Sugar}},\ }\bibfield  {title} {\bibinfo {title} {Monte carlo calculations of coupled boson-fermion systems. i},\ }\href {https://doi.org/10.1103/PhysRevD.24.2278} {\bibfield  {journal} {\bibinfo  {journal} {Phys. Rev. D}\ }\textbf {\bibinfo {volume} {24}},\ \bibinfo {pages} {2278} (\bibinfo {year} {1981})}\BibitemShut {NoStop}%
\bibitem [{\citenamefont {Hirsch}\ \emph {et~al.}(1982)\citenamefont {Hirsch}, \citenamefont {Sugar}, \citenamefont {Scalapino},\ and\ \citenamefont {Blankenbecler}}]{Hirsch82}%
  \BibitemOpen
  \bibfield  {author} {\bibinfo {author} {\bibfnamefont {J.~E.}\ \bibnamefont {Hirsch}}, \bibinfo {author} {\bibfnamefont {R.~L.}\ \bibnamefont {Sugar}}, \bibinfo {author} {\bibfnamefont {D.~J.}\ \bibnamefont {Scalapino}},\ and\ \bibinfo {author} {\bibfnamefont {R.}~\bibnamefont {Blankenbecler}},\ }\bibfield  {title} {\bibinfo {title} {Monte carlo simulations of one-dimensional fermion systems},\ }\href {https://doi.org/10.1103/PhysRevB.26.5033} {\bibfield  {journal} {\bibinfo  {journal} {Phys. Rev. B}\ }\textbf {\bibinfo {volume} {26}},\ \bibinfo {pages} {5033} (\bibinfo {year} {1982})}\BibitemShut {NoStop}%
\bibitem [{\citenamefont {Foulkes}\ \emph {et~al.}(2001)\citenamefont {Foulkes}, \citenamefont {Mitas}, \citenamefont {Needs},\ and\ \citenamefont {Rajagopal}}]{Foulkes01}%
  \BibitemOpen
  \bibfield  {author} {\bibinfo {author} {\bibfnamefont {W.~M.~C.}\ \bibnamefont {Foulkes}}, \bibinfo {author} {\bibfnamefont {L.}~\bibnamefont {Mitas}}, \bibinfo {author} {\bibfnamefont {R.~J.}\ \bibnamefont {Needs}},\ and\ \bibinfo {author} {\bibfnamefont {G.}~\bibnamefont {Rajagopal}},\ }\bibfield  {title} {\bibinfo {title} {Quantum monte carlo simulations of solids},\ }\href {https://doi.org/10.1103/RevModPhys.73.33} {\bibfield  {journal} {\bibinfo  {journal} {Rev. Mod. Phys.}\ }\textbf {\bibinfo {volume} {73}},\ \bibinfo {pages} {33} (\bibinfo {year} {2001})}\BibitemShut {NoStop}%
\bibitem [{\citenamefont {Schollw{\"o}ck}(2011)}]{Schollwock11}%
  \BibitemOpen
  \bibfield  {author} {\bibinfo {author} {\bibfnamefont {U.}~\bibnamefont {Schollw{\"o}ck}},\ }\bibfield  {title} {\bibinfo {title} {The density-matrix renormalization group in the age of matrix product states},\ }\href {https://doi.org/10.1016/j.aop.2010.09.012} {\bibfield  {journal} {\bibinfo  {journal} {Ann.~Phys.~(N.~Y.)}\ }\textbf {\bibinfo {volume} {326}},\ \bibinfo {pages} {96} (\bibinfo {year} {2011})}\BibitemShut {NoStop}%
\bibitem [{\citenamefont {Or{\'{u}}s}(2014)}]{Orus2014}%
  \BibitemOpen
  \bibfield  {author} {\bibinfo {author} {\bibfnamefont {R.}~\bibnamefont {Or{\'{u}}s}},\ }\bibfield  {title} {\bibinfo {title} {A practical introduction to tensor networks: Matrix product states and projected entangled pair states},\ }\href {https://doi.org/10.1016/j.aop.2014.06.013} {\bibfield  {journal} {\bibinfo  {journal} {Annals of Physics}\ }\textbf {\bibinfo {volume} {349}},\ \bibinfo {pages} {117} (\bibinfo {year} {2014})}\BibitemShut {NoStop}%
\bibitem [{\citenamefont {Troyer}\ and\ \citenamefont {Wiese}(2005)}]{Troyer05}%
  \BibitemOpen
  \bibfield  {author} {\bibinfo {author} {\bibfnamefont {M.}~\bibnamefont {Troyer}}\ and\ \bibinfo {author} {\bibfnamefont {U.-J.}\ \bibnamefont {Wiese}},\ }\bibfield  {title} {\bibinfo {title} {Computational complexity and fundamental limitations to fermionic quantum monte carlo simulations},\ }\href {https://doi.org/10.1103/PhysRevLett.94.170201} {\bibfield  {journal} {\bibinfo  {journal} {Phys. Rev. Lett.}\ }\textbf {\bibinfo {volume} {94}},\ \bibinfo {pages} {170201} (\bibinfo {year} {2005})}\BibitemShut {NoStop}%
\bibitem [{\citenamefont {Schuch}\ \emph {et~al.}(2007)\citenamefont {Schuch}, \citenamefont {Wolf}, \citenamefont {Verstraete},\ and\ \citenamefont {Cirac}}]{Schuch07}%
  \BibitemOpen
  \bibfield  {author} {\bibinfo {author} {\bibfnamefont {N.}~\bibnamefont {Schuch}}, \bibinfo {author} {\bibfnamefont {M.~M.}\ \bibnamefont {Wolf}}, \bibinfo {author} {\bibfnamefont {F.}~\bibnamefont {Verstraete}},\ and\ \bibinfo {author} {\bibfnamefont {J.~I.}\ \bibnamefont {Cirac}},\ }\bibfield  {title} {\bibinfo {title} {Computational complexity of projected entangled pair states},\ }\href {https://doi.org/10.1103/PhysRevLett.98.140506} {\bibfield  {journal} {\bibinfo  {journal} {Phys. Rev. Lett.}\ }\textbf {\bibinfo {volume} {98}},\ \bibinfo {pages} {140506} (\bibinfo {year} {2007})}\BibitemShut {NoStop}%
\bibitem [{\citenamefont {Haferkamp}\ \emph {et~al.}(2020)\citenamefont {Haferkamp}, \citenamefont {Hangleiter}, \citenamefont {Eisert},\ and\ \citenamefont {Gluza}}]{Haferkamp20}%
  \BibitemOpen
  \bibfield  {author} {\bibinfo {author} {\bibfnamefont {J.}~\bibnamefont {Haferkamp}}, \bibinfo {author} {\bibfnamefont {D.}~\bibnamefont {Hangleiter}}, \bibinfo {author} {\bibfnamefont {J.}~\bibnamefont {Eisert}},\ and\ \bibinfo {author} {\bibfnamefont {M.}~\bibnamefont {Gluza}},\ }\bibfield  {title} {\bibinfo {title} {Contracting projected entangled pair states is average-case hard},\ }\href {https://doi.org/10.1103/PhysRevResearch.2.013010} {\bibfield  {journal} {\bibinfo  {journal} {Phys. Rev. Res.}\ }\textbf {\bibinfo {volume} {2}},\ \bibinfo {pages} {013010} (\bibinfo {year} {2020})}\BibitemShut {NoStop}%
\bibitem [{\citenamefont {Bañuls}\ \emph {et~al.}(2020)\citenamefont {Bañuls}, \citenamefont {Blatt}, \citenamefont {Catani}, \citenamefont {Celi}, \citenamefont {Cirac}, \citenamefont {Dalmonte}, \citenamefont {Fallani}, \citenamefont {Jansen}, \citenamefont {Lewenstein}, \citenamefont {Montangero}, \citenamefont {Muschik}, \citenamefont {Reznik}, \citenamefont {Rico}, \citenamefont {Tagliacozzo}, \citenamefont {Van~Acoleyen}, \citenamefont {Verstraete}, \citenamefont {Wiese}, \citenamefont {Wingate}, \citenamefont {Zakrzewski},\ and\ \citenamefont {Zoller}}]{Banuls20}%
  \BibitemOpen
  \bibfield  {author} {\bibinfo {author} {\bibfnamefont {M.~C.}\ \bibnamefont {Bañuls}}, \bibinfo {author} {\bibfnamefont {R.}~\bibnamefont {Blatt}}, \bibinfo {author} {\bibfnamefont {J.}~\bibnamefont {Catani}}, \bibinfo {author} {\bibfnamefont {A.}~\bibnamefont {Celi}}, \bibinfo {author} {\bibfnamefont {J.~I.}\ \bibnamefont {Cirac}}, \bibinfo {author} {\bibfnamefont {M.}~\bibnamefont {Dalmonte}}, \bibinfo {author} {\bibfnamefont {L.}~\bibnamefont {Fallani}}, \bibinfo {author} {\bibfnamefont {K.}~\bibnamefont {Jansen}}, \bibinfo {author} {\bibfnamefont {M.}~\bibnamefont {Lewenstein}}, \bibinfo {author} {\bibfnamefont {S.}~\bibnamefont {Montangero}}, \bibinfo {author} {\bibfnamefont {C.~A.}\ \bibnamefont {Muschik}}, \bibinfo {author} {\bibfnamefont {B.}~\bibnamefont {Reznik}}, \bibinfo {author} {\bibfnamefont {E.}~\bibnamefont {Rico}}, \bibinfo {author} {\bibfnamefont {L.}~\bibnamefont {Tagliacozzo}}, \bibinfo {author} {\bibfnamefont {K.}~\bibnamefont {Van~Acoleyen}}, \bibinfo {author} {\bibfnamefont
  {F.}~\bibnamefont {Verstraete}}, \bibinfo {author} {\bibfnamefont {U.-J.}\ \bibnamefont {Wiese}}, \bibinfo {author} {\bibfnamefont {M.}~\bibnamefont {Wingate}}, \bibinfo {author} {\bibfnamefont {J.}~\bibnamefont {Zakrzewski}},\ and\ \bibinfo {author} {\bibfnamefont {P.}~\bibnamefont {Zoller}},\ }\bibfield  {title} {\bibinfo {title} {Simulating lattice gauge theories within quantum technologies},\ }\bibfield  {journal} {\bibinfo  {journal} {The European Physical Journal D}\ }\textbf {\bibinfo {volume} {74}},\ \href {https://doi.org/10.1140/epjd/e2020-100571-8} {10.1140/epjd/e2020-100571-8} (\bibinfo {year} {2020})\BibitemShut {NoStop}%
\bibitem [{\citenamefont {Bauer}\ \emph {et~al.}(2023)\citenamefont {Bauer}, \citenamefont {Davoudi}, \citenamefont {Balantekin}, \citenamefont {Bhattacharya}, \citenamefont {Carena}, \citenamefont {de~Jong}, \citenamefont {Draper}, \citenamefont {El-Khadra}, \citenamefont {Gemelke}, \citenamefont {Hanada}, \citenamefont {Kharzeev}, \citenamefont {Lamm}, \citenamefont {Li}, \citenamefont {Liu}, \citenamefont {Lukin}, \citenamefont {Meurice}, \citenamefont {Monroe}, \citenamefont {Nachman}, \citenamefont {Pagano}, \citenamefont {Preskill}, \citenamefont {Rinaldi}, \citenamefont {Roggero}, \citenamefont {Santiago}, \citenamefont {Savage}, \citenamefont {Siddiqi}, \citenamefont {Siopsis}, \citenamefont {Van~Zanten}, \citenamefont {Wiebe}, \citenamefont {Yamauchi}, \citenamefont {Yeter-Aydeniz},\ and\ \citenamefont {Zorzetti}}]{Bauer23}%
  \BibitemOpen
  \bibfield  {author} {\bibinfo {author} {\bibfnamefont {C.~W.}\ \bibnamefont {Bauer}}, \bibinfo {author} {\bibfnamefont {Z.}~\bibnamefont {Davoudi}}, \bibinfo {author} {\bibfnamefont {A.~B.}\ \bibnamefont {Balantekin}}, \bibinfo {author} {\bibfnamefont {T.}~\bibnamefont {Bhattacharya}}, \bibinfo {author} {\bibfnamefont {M.}~\bibnamefont {Carena}}, \bibinfo {author} {\bibfnamefont {W.~A.}\ \bibnamefont {de~Jong}}, \bibinfo {author} {\bibfnamefont {P.}~\bibnamefont {Draper}}, \bibinfo {author} {\bibfnamefont {A.}~\bibnamefont {El-Khadra}}, \bibinfo {author} {\bibfnamefont {N.}~\bibnamefont {Gemelke}}, \bibinfo {author} {\bibfnamefont {M.}~\bibnamefont {Hanada}}, \bibinfo {author} {\bibfnamefont {D.}~\bibnamefont {Kharzeev}}, \bibinfo {author} {\bibfnamefont {H.}~\bibnamefont {Lamm}}, \bibinfo {author} {\bibfnamefont {Y.-Y.}\ \bibnamefont {Li}}, \bibinfo {author} {\bibfnamefont {J.}~\bibnamefont {Liu}}, \bibinfo {author} {\bibfnamefont {M.}~\bibnamefont {Lukin}}, \bibinfo {author} {\bibfnamefont
  {Y.}~\bibnamefont {Meurice}}, \bibinfo {author} {\bibfnamefont {C.}~\bibnamefont {Monroe}}, \bibinfo {author} {\bibfnamefont {B.}~\bibnamefont {Nachman}}, \bibinfo {author} {\bibfnamefont {G.}~\bibnamefont {Pagano}}, \bibinfo {author} {\bibfnamefont {J.}~\bibnamefont {Preskill}}, \bibinfo {author} {\bibfnamefont {E.}~\bibnamefont {Rinaldi}}, \bibinfo {author} {\bibfnamefont {A.}~\bibnamefont {Roggero}}, \bibinfo {author} {\bibfnamefont {D.~I.}\ \bibnamefont {Santiago}}, \bibinfo {author} {\bibfnamefont {M.~J.}\ \bibnamefont {Savage}}, \bibinfo {author} {\bibfnamefont {I.}~\bibnamefont {Siddiqi}}, \bibinfo {author} {\bibfnamefont {G.}~\bibnamefont {Siopsis}}, \bibinfo {author} {\bibfnamefont {D.}~\bibnamefont {Van~Zanten}}, \bibinfo {author} {\bibfnamefont {N.}~\bibnamefont {Wiebe}}, \bibinfo {author} {\bibfnamefont {Y.}~\bibnamefont {Yamauchi}}, \bibinfo {author} {\bibfnamefont {K.}~\bibnamefont {Yeter-Aydeniz}},\ and\ \bibinfo {author} {\bibfnamefont {S.}~\bibnamefont {Zorzetti}},\ }\bibfield  {title}
  {\bibinfo {title} {Quantum simulation for high-energy physics},\ }\href {https://doi.org/10.1103/PRXQuantum.4.027001} {\bibfield  {journal} {\bibinfo  {journal} {PRX Quantum}\ }\textbf {\bibinfo {volume} {4}},\ \bibinfo {pages} {027001} (\bibinfo {year} {2023})}\BibitemShut {NoStop}%
\bibitem [{\citenamefont {Bernien}\ \emph {et~al.}(2017)\citenamefont {Bernien}, \citenamefont {Schwartz}, \citenamefont {Keesling}, \citenamefont {Levine}, \citenamefont {Omran}, \citenamefont {Pichler}, \citenamefont {Choi}, \citenamefont {Zibrov}, \citenamefont {Endres}, \citenamefont {Greiner} \emph {et~al.}}]{Bernien17}%
  \BibitemOpen
  \bibfield  {author} {\bibinfo {author} {\bibfnamefont {H.}~\bibnamefont {Bernien}}, \bibinfo {author} {\bibfnamefont {S.}~\bibnamefont {Schwartz}}, \bibinfo {author} {\bibfnamefont {A.}~\bibnamefont {Keesling}}, \bibinfo {author} {\bibfnamefont {H.}~\bibnamefont {Levine}}, \bibinfo {author} {\bibfnamefont {A.}~\bibnamefont {Omran}}, \bibinfo {author} {\bibfnamefont {H.}~\bibnamefont {Pichler}}, \bibinfo {author} {\bibfnamefont {S.}~\bibnamefont {Choi}}, \bibinfo {author} {\bibfnamefont {A.~S.}\ \bibnamefont {Zibrov}}, \bibinfo {author} {\bibfnamefont {M.}~\bibnamefont {Endres}}, \bibinfo {author} {\bibfnamefont {M.}~\bibnamefont {Greiner}}, \emph {et~al.},\ }\bibfield  {title} {\bibinfo {title} {Probing many-body dynamics on a 51-atom quantum simulator},\ }\href {https://doi.org/10.1038/nature24622} {\bibfield  {journal} {\bibinfo  {journal} {Nature}\ }\textbf {\bibinfo {volume} {551}},\ \bibinfo {pages} {579} (\bibinfo {year} {2017})}\BibitemShut {NoStop}%
\bibitem [{\citenamefont {Surace}\ \emph {et~al.}(2020)\citenamefont {Surace}, \citenamefont {Mazza}, \citenamefont {Giudici}, \citenamefont {Lerose}, \citenamefont {Gambassi},\ and\ \citenamefont {Dalmonte}}]{Surace19}%
  \BibitemOpen
  \bibfield  {author} {\bibinfo {author} {\bibfnamefont {F.~M.}\ \bibnamefont {Surace}}, \bibinfo {author} {\bibfnamefont {P.~P.}\ \bibnamefont {Mazza}}, \bibinfo {author} {\bibfnamefont {G.}~\bibnamefont {Giudici}}, \bibinfo {author} {\bibfnamefont {A.}~\bibnamefont {Lerose}}, \bibinfo {author} {\bibfnamefont {A.}~\bibnamefont {Gambassi}},\ and\ \bibinfo {author} {\bibfnamefont {M.}~\bibnamefont {Dalmonte}},\ }\bibfield  {title} {\bibinfo {title} {{Lattice Gauge Theories and String Dynamics in {Rydberg} Atom Quantum Simulators}},\ }\href {https://doi.org/10.1103/PhysRevX.10.021041} {\bibfield  {journal} {\bibinfo  {journal} {Phys. Rev. X}\ }\textbf {\bibinfo {volume} {10}},\ \bibinfo {pages} {021041} (\bibinfo {year} {2020})}\BibitemShut {NoStop}%
\bibitem [{\citenamefont {Yang}\ \emph {et~al.}(2020)\citenamefont {Yang}, \citenamefont {Liu}, \citenamefont {Gorshkov},\ and\ \citenamefont {Iadecola}}]{Yang20}%
  \BibitemOpen
  \bibfield  {author} {\bibinfo {author} {\bibfnamefont {Z.-C.}\ \bibnamefont {Yang}}, \bibinfo {author} {\bibfnamefont {F.}~\bibnamefont {Liu}}, \bibinfo {author} {\bibfnamefont {A.~V.}\ \bibnamefont {Gorshkov}},\ and\ \bibinfo {author} {\bibfnamefont {T.}~\bibnamefont {Iadecola}},\ }\bibfield  {title} {\bibinfo {title} {Hilbert-space fragmentation from strict confinement},\ }\href {https://doi.org/10.1103/PhysRevLett.124.207602} {\bibfield  {journal} {\bibinfo  {journal} {Phys. Rev. Lett.}\ }\textbf {\bibinfo {volume} {124}},\ \bibinfo {pages} {207602} (\bibinfo {year} {2020})}\BibitemShut {NoStop}%
\bibitem [{\citenamefont {Borla}\ \emph {et~al.}(2020)\citenamefont {Borla}, \citenamefont {Verresen}, \citenamefont {Grusdt},\ and\ \citenamefont {Moroz}}]{Borla20}%
  \BibitemOpen
  \bibfield  {author} {\bibinfo {author} {\bibfnamefont {U.}~\bibnamefont {Borla}}, \bibinfo {author} {\bibfnamefont {R.}~\bibnamefont {Verresen}}, \bibinfo {author} {\bibfnamefont {F.}~\bibnamefont {Grusdt}},\ and\ \bibinfo {author} {\bibfnamefont {S.}~\bibnamefont {Moroz}},\ }\bibfield  {title} {\bibinfo {title} {Confined phases of one-dimensional spinless fermions coupled to ${Z}_{2}$ gauge theory},\ }\href {https://doi.org/10.1103/PhysRevLett.124.120503} {\bibfield  {journal} {\bibinfo  {journal} {Phys. Rev. Lett.}\ }\textbf {\bibinfo {volume} {124}},\ \bibinfo {pages} {120503} (\bibinfo {year} {2020})}\BibitemShut {NoStop}%
\bibitem [{\citenamefont {Iadecola}\ and\ \citenamefont {Schecter}(2020)}]{Iadecola20}%
  \BibitemOpen
  \bibfield  {author} {\bibinfo {author} {\bibfnamefont {T.}~\bibnamefont {Iadecola}}\ and\ \bibinfo {author} {\bibfnamefont {M.}~\bibnamefont {Schecter}},\ }\bibfield  {title} {\bibinfo {title} {Quantum many-body scar states with emergent kinetic constraints and finite-entanglement revivals},\ }\href {https://doi.org/10.1103/PhysRevB.101.024306} {\bibfield  {journal} {\bibinfo  {journal} {Phys. Rev. B}\ }\textbf {\bibinfo {volume} {101}},\ \bibinfo {pages} {024306} (\bibinfo {year} {2020})}\BibitemShut {NoStop}%
\bibitem [{\citenamefont {Mildenberger}\ \emph {et~al.}(2022)\citenamefont {Mildenberger}, \citenamefont {Mruczkiewicz}, \citenamefont {Halimeh}, \citenamefont {Jiang},\ and\ \citenamefont {Hauke}}]{Mildenberger22}%
  \BibitemOpen
  \bibfield  {author} {\bibinfo {author} {\bibfnamefont {J.}~\bibnamefont {Mildenberger}}, \bibinfo {author} {\bibfnamefont {W.}~\bibnamefont {Mruczkiewicz}}, \bibinfo {author} {\bibfnamefont {J.~C.}\ \bibnamefont {Halimeh}}, \bibinfo {author} {\bibfnamefont {Z.}~\bibnamefont {Jiang}},\ and\ \bibinfo {author} {\bibfnamefont {P.}~\bibnamefont {Hauke}},\ }\href@noop {} {\bibinfo {title} {Probing confinement in a ${Z}_{2}$ lattice gauge theory on a quantum computer}} (\bibinfo {year} {2022}),\ \Eprint {https://arxiv.org/abs/arXiv:2203.08905} {arXiv:2203.08905} \BibitemShut {NoStop}%
\bibitem [{\citenamefont {Davoudi}\ \emph {et~al.}(2023)\citenamefont {Davoudi}, \citenamefont {Mueller},\ and\ \citenamefont {Powers}}]{Davoudi23}%
  \BibitemOpen
  \bibfield  {author} {\bibinfo {author} {\bibfnamefont {Z.}~\bibnamefont {Davoudi}}, \bibinfo {author} {\bibfnamefont {N.}~\bibnamefont {Mueller}},\ and\ \bibinfo {author} {\bibfnamefont {C.}~\bibnamefont {Powers}},\ }\bibfield  {title} {\bibinfo {title} {Towards quantum computing phase diagrams of gauge theories with thermal pure quantum states},\ }\href {https://doi.org/10.1103/PhysRevLett.131.081901} {\bibfield  {journal} {\bibinfo  {journal} {Phys. Rev. Lett.}\ }\textbf {\bibinfo {volume} {131}},\ \bibinfo {pages} {081901} (\bibinfo {year} {2023})}\BibitemShut {NoStop}%
\bibitem [{\citenamefont {Desaules}\ \emph {et~al.}(2024)\citenamefont {Desaules}, \citenamefont {Iadecola},\ and\ \citenamefont {Halimeh}}]{Desaules24}%
  \BibitemOpen
  \bibfield  {author} {\bibinfo {author} {\bibfnamefont {J.-Y.}\ \bibnamefont {Desaules}}, \bibinfo {author} {\bibfnamefont {T.}~\bibnamefont {Iadecola}},\ and\ \bibinfo {author} {\bibfnamefont {J.~C.}\ \bibnamefont {Halimeh}},\ }\href@noop {} {\bibinfo {title} {Mass-assisted local deconfinement in a confined $\mathbb{Z}_2$ lattice gauge theory}} (\bibinfo {year} {2024}),\ \Eprint {https://arxiv.org/abs/arXiv:2404.11645} {arXiv:2404.11645} \BibitemShut {NoStop}%
\bibitem [{\citenamefont {Kebrič}\ \emph {et~al.}(2024)\citenamefont {Kebrič}, \citenamefont {Halimeh}, \citenamefont {Schollwöck},\ and\ \citenamefont {Grusdt}}]{Kebrič24}%
  \BibitemOpen
  \bibfield  {author} {\bibinfo {author} {\bibfnamefont {M.}~\bibnamefont {Kebrič}}, \bibinfo {author} {\bibfnamefont {J.~C.}\ \bibnamefont {Halimeh}}, \bibinfo {author} {\bibfnamefont {U.}~\bibnamefont {Schollwöck}},\ and\ \bibinfo {author} {\bibfnamefont {F.}~\bibnamefont {Grusdt}},\ }\href {https://doi.org/10.48550/arXiv.2308.08592} {\bibinfo {title} {Confinement in 1+1d $\mathbb{Z}_2$ lattice gauge theories at finite temperature}} (\bibinfo {year} {2024})\BibitemShut {NoStop}%
\bibitem [{\citenamefont {Mark}\ \emph {et~al.}(2020)\citenamefont {Mark}, \citenamefont {Lin},\ and\ \citenamefont {Motrunich}}]{Mark20a}%
  \BibitemOpen
  \bibfield  {author} {\bibinfo {author} {\bibfnamefont {D.~K.}\ \bibnamefont {Mark}}, \bibinfo {author} {\bibfnamefont {C.-J.}\ \bibnamefont {Lin}},\ and\ \bibinfo {author} {\bibfnamefont {O.~I.}\ \bibnamefont {Motrunich}},\ }\bibfield  {title} {\bibinfo {title} {Unified structure for exact towers of scar states in the affleck-kennedy-lieb-tasaki and other models},\ }\href {https://doi.org/10.1103/PhysRevB.101.195131} {\bibfield  {journal} {\bibinfo  {journal} {Phys. Rev. B}\ }\textbf {\bibinfo {volume} {101}},\ \bibinfo {pages} {195131} (\bibinfo {year} {2020})}\BibitemShut {NoStop}%
\bibitem [{\citenamefont {Aramthottil}\ \emph {et~al.}(2022)\citenamefont {Aramthottil}, \citenamefont {Bhattacharya}, \citenamefont {Gonz\'alez-Cuadra}, \citenamefont {Lewenstein}, \citenamefont {Barbiero},\ and\ \citenamefont {Zakrzewski}}]{Aramthottil22}%
  \BibitemOpen
  \bibfield  {author} {\bibinfo {author} {\bibfnamefont {A.~S.}\ \bibnamefont {Aramthottil}}, \bibinfo {author} {\bibfnamefont {U.}~\bibnamefont {Bhattacharya}}, \bibinfo {author} {\bibfnamefont {D.}~\bibnamefont {Gonz\'alez-Cuadra}}, \bibinfo {author} {\bibfnamefont {M.}~\bibnamefont {Lewenstein}}, \bibinfo {author} {\bibfnamefont {L.}~\bibnamefont {Barbiero}},\ and\ \bibinfo {author} {\bibfnamefont {J.}~\bibnamefont {Zakrzewski}},\ }\bibfield  {title} {\bibinfo {title} {Scar states in deconfined ${Z}_{2}$ lattice gauge theories},\ }\href {https://doi.org/10.1103/PhysRevB.106.L041101} {\bibfield  {journal} {\bibinfo  {journal} {Phys. Rev. B}\ }\textbf {\bibinfo {volume} {106}},\ \bibinfo {pages} {L041101} (\bibinfo {year} {2022})}\BibitemShut {NoStop}%
\bibitem [{\citenamefont {Gustafson}\ \emph {et~al.}(2023)\citenamefont {Gustafson}, \citenamefont {Li}, \citenamefont {Khan}, \citenamefont {Kim}, \citenamefont {Kurkcuoglu}, \citenamefont {Alam}, \citenamefont {Orth}, \citenamefont {Rahmani},\ and\ \citenamefont {Iadecola}}]{Gustafson23}%
  \BibitemOpen
  \bibfield  {author} {\bibinfo {author} {\bibfnamefont {E.~J.}\ \bibnamefont {Gustafson}}, \bibinfo {author} {\bibfnamefont {A.~C.~Y.}\ \bibnamefont {Li}}, \bibinfo {author} {\bibfnamefont {A.}~\bibnamefont {Khan}}, \bibinfo {author} {\bibfnamefont {J.}~\bibnamefont {Kim}}, \bibinfo {author} {\bibfnamefont {D.~M.}\ \bibnamefont {Kurkcuoglu}}, \bibinfo {author} {\bibfnamefont {M.~S.}\ \bibnamefont {Alam}}, \bibinfo {author} {\bibfnamefont {P.~P.}\ \bibnamefont {Orth}}, \bibinfo {author} {\bibfnamefont {A.}~\bibnamefont {Rahmani}},\ and\ \bibinfo {author} {\bibfnamefont {T.}~\bibnamefont {Iadecola}},\ }\bibfield  {title} {\bibinfo {title} {Preparing quantum many-body scar states on quantum computers},\ }\href {https://doi.org/10.22331/q-2023-11-07-1171} {\bibfield  {journal} {\bibinfo  {journal} {Quantum}\ }\textbf {\bibinfo {volume} {7}},\ \bibinfo {pages} {1171} (\bibinfo {year} {2023})}\BibitemShut {NoStop}%
\bibitem [{\citenamefont {White}(2009)}]{White09}%
  \BibitemOpen
  \bibfield  {author} {\bibinfo {author} {\bibfnamefont {S.~R.}\ \bibnamefont {White}},\ }\bibfield  {title} {\bibinfo {title} {Minimally entangled typical quantum states at finite temperature},\ }\href {https://doi.org/10.1103/PhysRevLett.102.190601} {\bibfield  {journal} {\bibinfo  {journal} {Phys. Rev. Lett.}\ }\textbf {\bibinfo {volume} {102}},\ \bibinfo {pages} {190601} (\bibinfo {year} {2009})}\BibitemShut {NoStop}%
\bibitem [{\citenamefont {Stoudenmire}\ and\ \citenamefont {White}(2010)}]{stoudenmire2010}%
  \BibitemOpen
  \bibfield  {author} {\bibinfo {author} {\bibfnamefont {E.~M.}\ \bibnamefont {Stoudenmire}}\ and\ \bibinfo {author} {\bibfnamefont {S.~R.}\ \bibnamefont {White}},\ }\bibfield  {title} {\bibinfo {title} {Minimally entangled typical thermal state algorithms},\ }\href {https://doi.org/10.1088/1367-2630/12/5/055026} {\bibfield  {journal} {\bibinfo  {journal} {New Journal of Physics}\ }\textbf {\bibinfo {volume} {12}},\ \bibinfo {pages} {055026} (\bibinfo {year} {2010})}\BibitemShut {NoStop}%
\bibitem [{\citenamefont {Binder}\ and\ \citenamefont {Barthel}(2015)}]{Binder15}%
  \BibitemOpen
  \bibfield  {author} {\bibinfo {author} {\bibfnamefont {M.}~\bibnamefont {Binder}}\ and\ \bibinfo {author} {\bibfnamefont {T.}~\bibnamefont {Barthel}},\ }\bibfield  {title} {\bibinfo {title} {Minimally entangled typical thermal states versus matrix product purifications for the simulation of equilibrium states and time evolution},\ }\href {https://doi.org/10.1103/PhysRevB.92.125119} {\bibfield  {journal} {\bibinfo  {journal} {Phys. Rev. B}\ }\textbf {\bibinfo {volume} {92}},\ \bibinfo {pages} {125119} (\bibinfo {year} {2015})}\BibitemShut {NoStop}%
\bibitem [{\citenamefont {Binder}\ and\ \citenamefont {Barthel}(2017)}]{Binder17}%
  \BibitemOpen
  \bibfield  {author} {\bibinfo {author} {\bibfnamefont {M.}~\bibnamefont {Binder}}\ and\ \bibinfo {author} {\bibfnamefont {T.}~\bibnamefont {Barthel}},\ }\bibfield  {title} {\bibinfo {title} {Symmetric minimally entangled typical thermal states for canonical and grand-canonical ensembles},\ }\href {https://doi.org/10.1103/PhysRevB.95.195148} {\bibfield  {journal} {\bibinfo  {journal} {Phys. Rev. B}\ }\textbf {\bibinfo {volume} {95}},\ \bibinfo {pages} {195148} (\bibinfo {year} {2017})}\BibitemShut {NoStop}%
\bibitem [{\citenamefont {Wietek}\ \emph {et~al.}(2021{\natexlab{a}})\citenamefont {Wietek}, \citenamefont {He}, \citenamefont {White}, \citenamefont {Georges},\ and\ \citenamefont {Stoudenmire}}]{Wietek21a}%
  \BibitemOpen
  \bibfield  {author} {\bibinfo {author} {\bibfnamefont {A.}~\bibnamefont {Wietek}}, \bibinfo {author} {\bibfnamefont {Y.-Y.}\ \bibnamefont {He}}, \bibinfo {author} {\bibfnamefont {S.~R.}\ \bibnamefont {White}}, \bibinfo {author} {\bibfnamefont {A.}~\bibnamefont {Georges}},\ and\ \bibinfo {author} {\bibfnamefont {E.~M.}\ \bibnamefont {Stoudenmire}},\ }\bibfield  {title} {\bibinfo {title} {Stripes, antiferromagnetism, and the pseudogap in the doped hubbard model at finite temperature},\ }\href {https://doi.org/10.1103/PhysRevX.11.031007} {\bibfield  {journal} {\bibinfo  {journal} {Phys. Rev. X}\ }\textbf {\bibinfo {volume} {11}},\ \bibinfo {pages} {031007} (\bibinfo {year} {2021}{\natexlab{a}})}\BibitemShut {NoStop}%
\bibitem [{\citenamefont {Wietek}\ \emph {et~al.}(2021{\natexlab{b}})\citenamefont {Wietek}, \citenamefont {Rossi}, \citenamefont {\ifmmode~\check{S}\else \v{S}\fi{}imkovic}, \citenamefont {Klett}, \citenamefont {Hansmann}, \citenamefont {Ferrero}, \citenamefont {Stoudenmire}, \citenamefont {Sch\"afer},\ and\ \citenamefont {Georges}}]{Wietek21b}%
  \BibitemOpen
  \bibfield  {author} {\bibinfo {author} {\bibfnamefont {A.}~\bibnamefont {Wietek}}, \bibinfo {author} {\bibfnamefont {R.}~\bibnamefont {Rossi}}, \bibinfo {author} {\bibfnamefont {F.}~\bibnamefont {\ifmmode~\check{S}\else \v{S}\fi{}imkovic}}, \bibinfo {author} {\bibfnamefont {M.}~\bibnamefont {Klett}}, \bibinfo {author} {\bibfnamefont {P.}~\bibnamefont {Hansmann}}, \bibinfo {author} {\bibfnamefont {M.}~\bibnamefont {Ferrero}}, \bibinfo {author} {\bibfnamefont {E.~M.}\ \bibnamefont {Stoudenmire}}, \bibinfo {author} {\bibfnamefont {T.}~\bibnamefont {Sch\"afer}},\ and\ \bibinfo {author} {\bibfnamefont {A.}~\bibnamefont {Georges}},\ }\bibfield  {title} {\bibinfo {title} {Mott insulating states with competing orders in the triangular lattice hubbard model},\ }\href {https://doi.org/10.1103/PhysRevX.11.041013} {\bibfield  {journal} {\bibinfo  {journal} {Phys. Rev. X}\ }\textbf {\bibinfo {volume} {11}},\ \bibinfo {pages} {041013} (\bibinfo {year} {2021}{\natexlab{b}})}\BibitemShut {NoStop}%
\bibitem [{\citenamefont {Motta}\ \emph {et~al.}(2019)\citenamefont {Motta}, \citenamefont {Sun}, \citenamefont {Tan}, \citenamefont {O’Rourke}, \citenamefont {Ye}, \citenamefont {Minnich}, \citenamefont {Brandão},\ and\ \citenamefont {Chan}}]{Motta19}%
  \BibitemOpen
  \bibfield  {author} {\bibinfo {author} {\bibfnamefont {M.}~\bibnamefont {Motta}}, \bibinfo {author} {\bibfnamefont {C.}~\bibnamefont {Sun}}, \bibinfo {author} {\bibfnamefont {A.~T.~K.}\ \bibnamefont {Tan}}, \bibinfo {author} {\bibfnamefont {M.~J.}\ \bibnamefont {O’Rourke}}, \bibinfo {author} {\bibfnamefont {E.}~\bibnamefont {Ye}}, \bibinfo {author} {\bibfnamefont {A.~J.}\ \bibnamefont {Minnich}}, \bibinfo {author} {\bibfnamefont {F.~G. S.~L.}\ \bibnamefont {Brandão}},\ and\ \bibinfo {author} {\bibfnamefont {G.~K.-L.}\ \bibnamefont {Chan}},\ }\bibfield  {title} {\bibinfo {title} {Determining eigenstates and thermal states on a quantum computer using quantum imaginary time evolution},\ }\href {https://doi.org/10.1038/s41567-019-0704-4} {\bibfield  {journal} {\bibinfo  {journal} {Nature Physics}\ }\textbf {\bibinfo {volume} {16}},\ \bibinfo {pages} {205–210} (\bibinfo {year} {2019})}\BibitemShut {NoStop}%
\bibitem [{\citenamefont {Gomes}\ \emph {et~al.}(2021)\citenamefont {Gomes}, \citenamefont {Mukherjee}, \citenamefont {Zhang}, \citenamefont {Iadecola}, \citenamefont {Wang}, \citenamefont {Ho}, \citenamefont {Orth},\ and\ \citenamefont {Yao}}]{AVQITE}%
  \BibitemOpen
  \bibfield  {author} {\bibinfo {author} {\bibfnamefont {N.}~\bibnamefont {Gomes}}, \bibinfo {author} {\bibfnamefont {A.}~\bibnamefont {Mukherjee}}, \bibinfo {author} {\bibfnamefont {F.}~\bibnamefont {Zhang}}, \bibinfo {author} {\bibfnamefont {T.}~\bibnamefont {Iadecola}}, \bibinfo {author} {\bibfnamefont {C.-Z.}\ \bibnamefont {Wang}}, \bibinfo {author} {\bibfnamefont {K.-M.}\ \bibnamefont {Ho}}, \bibinfo {author} {\bibfnamefont {P.~P.}\ \bibnamefont {Orth}},\ and\ \bibinfo {author} {\bibfnamefont {Y.-X.}\ \bibnamefont {Yao}},\ }\bibfield  {title} {\bibinfo {title} {Adaptive variational quantum imaginary time evolution approach for ground state preparation},\ }\href {https://doi.org/10.1002/qute.202100114} {\bibfield  {journal} {\bibinfo  {journal} {Adv. Quantum Technol.}\ }\textbf {\bibinfo {volume} {4}},\ \bibinfo {pages} {2100114} (\bibinfo {year} {2021})}\BibitemShut {NoStop}%
\bibitem [{\citenamefont {McArdle}\ \emph {et~al.}(2019)\citenamefont {McArdle}, \citenamefont {Jones}, \citenamefont {Endo}, \citenamefont {Li}, \citenamefont {Benjamin},\ and\ \citenamefont {Yuan}}]{McArdle19}%
  \BibitemOpen
  \bibfield  {author} {\bibinfo {author} {\bibfnamefont {S.}~\bibnamefont {McArdle}}, \bibinfo {author} {\bibfnamefont {T.}~\bibnamefont {Jones}}, \bibinfo {author} {\bibfnamefont {S.}~\bibnamefont {Endo}}, \bibinfo {author} {\bibfnamefont {Y.}~\bibnamefont {Li}}, \bibinfo {author} {\bibfnamefont {S.~C.}\ \bibnamefont {Benjamin}},\ and\ \bibinfo {author} {\bibfnamefont {X.}~\bibnamefont {Yuan}},\ }\bibfield  {title} {\bibinfo {title} {Variational ansatz-based quantum simulation of imaginary time evolution},\ }\bibfield  {journal} {\bibinfo  {journal} {npj Quantum Information}\ }\textbf {\bibinfo {volume} {5}},\ \href {https://doi.org/10.1038/s41534-019-0187-2} {10.1038/s41534-019-0187-2} (\bibinfo {year} {2019})\BibitemShut {NoStop}%
\bibitem [{\citenamefont {Getelina}\ \emph {et~al.}(2023)\citenamefont {Getelina}, \citenamefont {Gomes}, \citenamefont {Iadecola}, \citenamefont {Orth},\ and\ \citenamefont {Yao}}]{Getelina23}%
  \BibitemOpen
  \bibfield  {author} {\bibinfo {author} {\bibfnamefont {J.~C.}\ \bibnamefont {Getelina}}, \bibinfo {author} {\bibfnamefont {N.}~\bibnamefont {Gomes}}, \bibinfo {author} {\bibfnamefont {T.}~\bibnamefont {Iadecola}}, \bibinfo {author} {\bibfnamefont {P.~P.}\ \bibnamefont {Orth}},\ and\ \bibinfo {author} {\bibfnamefont {Y.-X.}\ \bibnamefont {Yao}},\ }\bibfield  {title} {\bibinfo {title} {{Adaptive variational quantum minimally entangled typical thermal states for finite temperature simulations}},\ }\href {https://doi.org/10.21468/SciPostPhys.15.3.102} {\bibfield  {journal} {\bibinfo  {journal} {SciPost Phys.}\ }\textbf {\bibinfo {volume} {15}},\ \bibinfo {pages} {102} (\bibinfo {year} {2023})}\BibitemShut {NoStop}%
\bibitem [{\citenamefont {Fishman}\ \emph {et~al.}(2022)\citenamefont {Fishman}, \citenamefont {White},\ and\ \citenamefont {Stoudenmire}}]{itensor}%
  \BibitemOpen
  \bibfield  {author} {\bibinfo {author} {\bibfnamefont {M.}~\bibnamefont {Fishman}}, \bibinfo {author} {\bibfnamefont {S.~R.}\ \bibnamefont {White}},\ and\ \bibinfo {author} {\bibfnamefont {E.~M.}\ \bibnamefont {Stoudenmire}},\ }\bibfield  {title} {\bibinfo {title} {{The ITensor Software Library for Tensor Network Calculations}},\ }\href {https://doi.org/10.21468/SciPostPhysCodeb.4} {\bibfield  {journal} {\bibinfo  {journal} {SciPost Phys. Codebases}\ ,\ \bibinfo {pages} {4}} (\bibinfo {year} {2022})}\BibitemShut {NoStop}%
\bibitem [{\citenamefont {Yao}\ \emph {et~al.}(2024)\citenamefont {Yao}, \citenamefont {Getelina}, \citenamefont {Mukherjee}, \citenamefont {Gomes}, \citenamefont {Iadecola},\ and\ \citenamefont {Orth}}]{CyQC}%
  \BibitemOpen
  \bibfield  {author} {\bibinfo {author} {\bibfnamefont {Y.-X.}\ \bibnamefont {Yao}}, \bibinfo {author} {\bibfnamefont {J.~C.}\ \bibnamefont {Getelina}}, \bibinfo {author} {\bibfnamefont {A.}~\bibnamefont {Mukherjee}}, \bibinfo {author} {\bibfnamefont {N.}~\bibnamefont {Gomes}}, \bibinfo {author} {\bibfnamefont {T.}~\bibnamefont {Iadecola}},\ and\ \bibinfo {author} {\bibfnamefont {P.~P.}\ \bibnamefont {Orth}},\ }\href {https://doi.org/10.6084/m9.figshare.26298763.v2} {\bibinfo {title} {{CyQC: Quantum computing toolset for correlated materials simulations}}} (\bibinfo {year} {2024})\BibitemShut {NoStop}%
\bibitem [{\citenamefont {Chen}\ \emph {et~al.}(2023)\citenamefont {Chen}, \citenamefont {Pollock}, \citenamefont {Yao}, \citenamefont {Orth},\ and\ \citenamefont {Iadecola}}]{Chen23}%
  \BibitemOpen
  \bibfield  {author} {\bibinfo {author} {\bibfnamefont {I.-C.}\ \bibnamefont {Chen}}, \bibinfo {author} {\bibfnamefont {K.}~\bibnamefont {Pollock}}, \bibinfo {author} {\bibfnamefont {Y.-X.}\ \bibnamefont {Yao}}, \bibinfo {author} {\bibfnamefont {P.~P.}\ \bibnamefont {Orth}},\ and\ \bibinfo {author} {\bibfnamefont {T.}~\bibnamefont {Iadecola}},\ }\href {https://doi.org/10.48550/arXiv.2310.03924} {\bibinfo {title} {Problem-tailored simulation of energy transport on noisy quantum computers}} (\bibinfo {year} {2023})\BibitemShut {NoStop}%
\bibitem [{\citenamefont {Zwolak}\ and\ \citenamefont {Vidal}(2004)}]{Zwolak04}%
  \BibitemOpen
  \bibfield  {author} {\bibinfo {author} {\bibfnamefont {M.}~\bibnamefont {Zwolak}}\ and\ \bibinfo {author} {\bibfnamefont {G.}~\bibnamefont {Vidal}},\ }\bibfield  {title} {\bibinfo {title} {Mixed-state dynamics in one-dimensional quantum lattice systems: A time-dependent superoperator renormalization algorithm},\ }\href {https://doi.org/10.1103/PhysRevLett.93.207205} {\bibfield  {journal} {\bibinfo  {journal} {Phys. Rev. Lett.}\ }\textbf {\bibinfo {volume} {93}},\ \bibinfo {pages} {207205} (\bibinfo {year} {2004})}\BibitemShut {NoStop}%
\bibitem [{\citenamefont {Verstraete}\ \emph {et~al.}(2004)\citenamefont {Verstraete}, \citenamefont {Garc\'{\i}a-Ripoll},\ and\ \citenamefont {Cirac}}]{Verstraete04}%
  \BibitemOpen
  \bibfield  {author} {\bibinfo {author} {\bibfnamefont {F.}~\bibnamefont {Verstraete}}, \bibinfo {author} {\bibfnamefont {J.~J.}\ \bibnamefont {Garc\'{\i}a-Ripoll}},\ and\ \bibinfo {author} {\bibfnamefont {J.~I.}\ \bibnamefont {Cirac}},\ }\bibfield  {title} {\bibinfo {title} {Matrix product density operators: Simulation of finite-temperature and dissipative systems},\ }\href {https://doi.org/10.1103/PhysRevLett.93.207204} {\bibfield  {journal} {\bibinfo  {journal} {Phys. Rev. Lett.}\ }\textbf {\bibinfo {volume} {93}},\ \bibinfo {pages} {207204} (\bibinfo {year} {2004})}\BibitemShut {NoStop}%
\bibitem [{\citenamefont {Feiguin}\ and\ \citenamefont {White}(2005)}]{Feiguin05}%
  \BibitemOpen
  \bibfield  {author} {\bibinfo {author} {\bibfnamefont {A.~E.}\ \bibnamefont {Feiguin}}\ and\ \bibinfo {author} {\bibfnamefont {S.~R.}\ \bibnamefont {White}},\ }\bibfield  {title} {\bibinfo {title} {Finite-temperature density matrix renormalization using an enlarged hilbert space},\ }\href {https://doi.org/10.1103/PhysRevB.72.220401} {\bibfield  {journal} {\bibinfo  {journal} {Phys. Rev. B}\ }\textbf {\bibinfo {volume} {72}},\ \bibinfo {pages} {220401} (\bibinfo {year} {2005})}\BibitemShut {NoStop}%
\bibitem [{\citenamefont {Haegeman}\ \emph {et~al.}(2011)\citenamefont {Haegeman}, \citenamefont {Cirac}, \citenamefont {Osborne}, \citenamefont {Pi\ifmmode~\check{z}\else \v{z}\fi{}orn}, \citenamefont {Verschelde},\ and\ \citenamefont {Verstraete}}]{Haegeman11}%
  \BibitemOpen
  \bibfield  {author} {\bibinfo {author} {\bibfnamefont {J.}~\bibnamefont {Haegeman}}, \bibinfo {author} {\bibfnamefont {J.~I.}\ \bibnamefont {Cirac}}, \bibinfo {author} {\bibfnamefont {T.~J.}\ \bibnamefont {Osborne}}, \bibinfo {author} {\bibfnamefont {I.}~\bibnamefont {Pi\ifmmode~\check{z}\else \v{z}\fi{}orn}}, \bibinfo {author} {\bibfnamefont {H.}~\bibnamefont {Verschelde}},\ and\ \bibinfo {author} {\bibfnamefont {F.}~\bibnamefont {Verstraete}},\ }\bibfield  {title} {\bibinfo {title} {Time-dependent variational principle for quantum lattices},\ }\href {https://doi.org/10.1103/PhysRevLett.107.070601} {\bibfield  {journal} {\bibinfo  {journal} {Phys. Rev. Lett.}\ }\textbf {\bibinfo {volume} {107}},\ \bibinfo {pages} {070601} (\bibinfo {year} {2011})}\BibitemShut {NoStop}%
\bibitem [{\citenamefont {Haegeman}\ \emph {et~al.}(2016)\citenamefont {Haegeman}, \citenamefont {Lubich}, \citenamefont {Oseledets}, \citenamefont {Vandereycken},\ and\ \citenamefont {Verstraete}}]{Haegeman16}%
  \BibitemOpen
  \bibfield  {author} {\bibinfo {author} {\bibfnamefont {J.}~\bibnamefont {Haegeman}}, \bibinfo {author} {\bibfnamefont {C.}~\bibnamefont {Lubich}}, \bibinfo {author} {\bibfnamefont {I.}~\bibnamefont {Oseledets}}, \bibinfo {author} {\bibfnamefont {B.}~\bibnamefont {Vandereycken}},\ and\ \bibinfo {author} {\bibfnamefont {F.}~\bibnamefont {Verstraete}},\ }\bibfield  {title} {\bibinfo {title} {Unifying time evolution and optimization with matrix product states},\ }\href {https://doi.org/10.1103/PhysRevB.94.165116} {\bibfield  {journal} {\bibinfo  {journal} {Phys. Rev. B}\ }\textbf {\bibinfo {volume} {94}},\ \bibinfo {pages} {165116} (\bibinfo {year} {2016})}\BibitemShut {NoStop}%
\bibitem [{\citenamefont {Yuan}\ \emph {et~al.}(2019)\citenamefont {Yuan}, \citenamefont {Endo}, \citenamefont {Zhao}, \citenamefont {Li},\ and\ \citenamefont {Benjamin}}]{Yuan19}%
  \BibitemOpen
  \bibfield  {author} {\bibinfo {author} {\bibfnamefont {X.}~\bibnamefont {Yuan}}, \bibinfo {author} {\bibfnamefont {S.}~\bibnamefont {Endo}}, \bibinfo {author} {\bibfnamefont {Q.}~\bibnamefont {Zhao}}, \bibinfo {author} {\bibfnamefont {Y.}~\bibnamefont {Li}},\ and\ \bibinfo {author} {\bibfnamefont {S.~C.}\ \bibnamefont {Benjamin}},\ }\bibfield  {title} {\bibinfo {title} {Theory of variational quantum simulation},\ }\href {https://doi.org/10.22331/q-2019-10-07-191} {\bibfield  {journal} {\bibinfo  {journal} {Quantum}\ }\textbf {\bibinfo {volume} {3}},\ \bibinfo {pages} {191} (\bibinfo {year} {2019})}\BibitemShut {NoStop}%
\bibitem [{\citenamefont {Yao}\ \emph {et~al.}(2021)\citenamefont {Yao}, \citenamefont {Gomes}, \citenamefont {Zhang}, \citenamefont {Wang}, \citenamefont {Ho}, \citenamefont {Iadecola},\ and\ \citenamefont {Orth}}]{Yao-AVQDS-PRX_Q-2021}%
  \BibitemOpen
  \bibfield  {author} {\bibinfo {author} {\bibfnamefont {Y.-X.}\ \bibnamefont {Yao}}, \bibinfo {author} {\bibfnamefont {N.}~\bibnamefont {Gomes}}, \bibinfo {author} {\bibfnamefont {F.}~\bibnamefont {Zhang}}, \bibinfo {author} {\bibfnamefont {C.-Z.}\ \bibnamefont {Wang}}, \bibinfo {author} {\bibfnamefont {K.-M.}\ \bibnamefont {Ho}}, \bibinfo {author} {\bibfnamefont {T.}~\bibnamefont {Iadecola}},\ and\ \bibinfo {author} {\bibfnamefont {P.~P.}\ \bibnamefont {Orth}},\ }\bibfield  {title} {\bibinfo {title} {Adaptive variational quantum dynamics simulations},\ }\href {https://doi.org/10.1103/PRXQuantum.2.030307} {\bibfield  {journal} {\bibinfo  {journal} {PRX Quantum}\ }\textbf {\bibinfo {volume} {2}},\ \bibinfo {pages} {030307} (\bibinfo {year} {2021})}\BibitemShut {NoStop}%
\bibitem [{\citenamefont {Getelina}\ \emph {et~al.}(2024)\citenamefont {Getelina}, \citenamefont {Wang}, \citenamefont {Iadecola}, \citenamefont {Yao},\ and\ \citenamefont {Orth}}]{Getelina24}%
  \BibitemOpen
  \bibfield  {author} {\bibinfo {author} {\bibfnamefont {J.~C.}\ \bibnamefont {Getelina}}, \bibinfo {author} {\bibfnamefont {C.-Z.}\ \bibnamefont {Wang}}, \bibinfo {author} {\bibfnamefont {T.}~\bibnamefont {Iadecola}}, \bibinfo {author} {\bibfnamefont {Y.-X.}\ \bibnamefont {Yao}},\ and\ \bibinfo {author} {\bibfnamefont {P.~P.}\ \bibnamefont {Orth}},\ }\bibfield  {title} {\bibinfo {title} {{Adaptive variational ground state preparation for spin-1 models on qubit-based architectures}},\ }\href {https://doi.org/10.1103/PhysRevB.109.085128} {\bibfield  {journal} {\bibinfo  {journal} {Phys. Rev. B}\ }\textbf {\bibinfo {volume} {109}},\ \bibinfo {pages} {085128} (\bibinfo {year} {2024})}\BibitemShut {NoStop}%
\bibitem [{\citenamefont {Khindanov}\ \emph {et~al.}()\citenamefont {Khindanov}, \citenamefont {Iadecola},\ and\ \citenamefont {Yao}}]{Khindanov2025inpreparation}%
  \BibitemOpen
  \bibfield  {author} {\bibinfo {author} {\bibfnamefont {A.}~\bibnamefont {Khindanov}}, \bibinfo {author} {\bibfnamefont {T.}~\bibnamefont {Iadecola}},\ and\ \bibinfo {author} {\bibfnamefont {Y.}~\bibnamefont {Yao}},\ }\bibfield  {title} {\bibinfo {title} {Tensor network simulations of adaptive variational quantum algorithms},\ }\href@noop {} {\bibinfo  {journal} {in preparation (2025)}\ }\BibitemShut {NoStop}%
\bibitem [{\citenamefont {Gray}(2018)}]{gray2018quimb}%
  \BibitemOpen
\bibfield  {journal} {  }\bibfield  {author} {\bibinfo {author} {\bibfnamefont {J.}~\bibnamefont {Gray}},\ }\bibfield  {title} {\bibinfo {title} {quimb: a python library for quantum information and many-body calculations},\ }\href {https://doi.org/10.21105/joss.00819} {\bibfield  {journal} {\bibinfo  {journal} {Journal of Open Source Software}\ }\textbf {\bibinfo {volume} {3}},\ \bibinfo {pages} {819} (\bibinfo {year} {2018})}\BibitemShut {NoStop}%
\bibitem [{\citenamefont {Gray}\ and\ \citenamefont {Kourtis}(2021)}]{Gray2021hyperoptimized}%
  \BibitemOpen
  \bibfield  {author} {\bibinfo {author} {\bibfnamefont {J.}~\bibnamefont {Gray}}\ and\ \bibinfo {author} {\bibfnamefont {S.}~\bibnamefont {Kourtis}},\ }\bibfield  {title} {\bibinfo {title} {Hyper-optimized tensor network contraction},\ }\href {https://doi.org/10.22331/q-2021-03-15-410} {\bibfield  {journal} {\bibinfo  {journal} {{Quantum}}\ }\textbf {\bibinfo {volume} {5}},\ \bibinfo {pages} {410} (\bibinfo {year} {2021})}\BibitemShut {NoStop}%
\bibitem [{\citenamefont {Zimborás}\ \emph {et~al.}(2025)\citenamefont {Zimborás}, \citenamefont {Koczor}, \citenamefont {Holmes}, \citenamefont {Borrelli}, \citenamefont {Gilyén}, \citenamefont {Huang}, \citenamefont {Cai}, \citenamefont {Acín}, \citenamefont {Aolita}, \citenamefont {Banchi}, \citenamefont {Brandão}, \citenamefont {Cavalcanti}, \citenamefont {Cubitt}, \citenamefont {Filippov}, \citenamefont {García-Pérez}, \citenamefont {Goold}, \citenamefont {Kálmán}, \citenamefont {Kyoseva}, \citenamefont {Rossi}, \citenamefont {Sokolov}, \citenamefont {Tavernelli},\ and\ \citenamefont {Maniscalco}}]{Zimboras2025}%
  \BibitemOpen
  \bibfield  {author} {\bibinfo {author} {\bibfnamefont {Z.}~\bibnamefont {Zimborás}}, \bibinfo {author} {\bibfnamefont {B.}~\bibnamefont {Koczor}}, \bibinfo {author} {\bibfnamefont {Z.}~\bibnamefont {Holmes}}, \bibinfo {author} {\bibfnamefont {E.-M.}\ \bibnamefont {Borrelli}}, \bibinfo {author} {\bibfnamefont {A.}~\bibnamefont {Gilyén}}, \bibinfo {author} {\bibfnamefont {H.-Y.}\ \bibnamefont {Huang}}, \bibinfo {author} {\bibfnamefont {Z.}~\bibnamefont {Cai}}, \bibinfo {author} {\bibfnamefont {A.}~\bibnamefont {Acín}}, \bibinfo {author} {\bibfnamefont {L.}~\bibnamefont {Aolita}}, \bibinfo {author} {\bibfnamefont {L.}~\bibnamefont {Banchi}}, \bibinfo {author} {\bibfnamefont {F.~G. S.~L.}\ \bibnamefont {Brandão}}, \bibinfo {author} {\bibfnamefont {D.}~\bibnamefont {Cavalcanti}}, \bibinfo {author} {\bibfnamefont {T.}~\bibnamefont {Cubitt}}, \bibinfo {author} {\bibfnamefont {S.~N.}\ \bibnamefont {Filippov}}, \bibinfo {author} {\bibfnamefont {G.}~\bibnamefont {García-Pérez}}, \bibinfo {author} {\bibfnamefont
  {J.}~\bibnamefont {Goold}}, \bibinfo {author} {\bibfnamefont {O.}~\bibnamefont {Kálmán}}, \bibinfo {author} {\bibfnamefont {E.}~\bibnamefont {Kyoseva}}, \bibinfo {author} {\bibfnamefont {M.~A.~C.}\ \bibnamefont {Rossi}}, \bibinfo {author} {\bibfnamefont {B.}~\bibnamefont {Sokolov}}, \bibinfo {author} {\bibfnamefont {I.}~\bibnamefont {Tavernelli}},\ and\ \bibinfo {author} {\bibfnamefont {S.}~\bibnamefont {Maniscalco}},\ }\href {https://arxiv.org/abs/2501.05694} {\bibinfo {title} {Myths around quantum computation before full fault tolerance: What no-go theorems rule out and what they don't}} (\bibinfo {year} {2025}),\ \Eprint {https://arxiv.org/abs/2501.05694} {arXiv:2501.05694 [quant-ph]} \BibitemShut {NoStop}%
\bibitem [{\citenamefont {Zhang}\ \emph {et~al.}(2025)\citenamefont {Zhang}, \citenamefont {Wang}, \citenamefont {Iadecola}, \citenamefont {Orth},\ and\ \citenamefont {Yao}}]{Zhang2025}%
  \BibitemOpen
  \bibfield  {author} {\bibinfo {author} {\bibfnamefont {F.}~\bibnamefont {Zhang}}, \bibinfo {author} {\bibfnamefont {C.-Z.}\ \bibnamefont {Wang}}, \bibinfo {author} {\bibfnamefont {T.}~\bibnamefont {Iadecola}}, \bibinfo {author} {\bibfnamefont {P.~P.}\ \bibnamefont {Orth}},\ and\ \bibinfo {author} {\bibfnamefont {Y.-X.}\ \bibnamefont {Yao}},\ }\bibfield  {title} {\bibinfo {title} {Adaptive variational quantum dynamics simulations with compressed circuits and fewer measurements},\ }\href {https://doi.org/10.1103/PhysRevB.111.094310} {\bibfield  {journal} {\bibinfo  {journal} {Phys. Rev. B}\ }\textbf {\bibinfo {volume} {111}},\ \bibinfo {pages} {094310} (\bibinfo {year} {2025})}\BibitemShut {NoStop}%
\bibitem [{\citenamefont {Shtanko}\ and\ \citenamefont {Movassagh}(2023)}]{Shtanko23}%
  \BibitemOpen
  \bibfield  {author} {\bibinfo {author} {\bibfnamefont {O.}~\bibnamefont {Shtanko}}\ and\ \bibinfo {author} {\bibfnamefont {R.}~\bibnamefont {Movassagh}},\ }\href {https://arxiv.org/abs/2112.14688} {\bibinfo {title} {Preparing thermal states on noiseless and noisy programmable quantum processors}} (\bibinfo {year} {2023}),\ \Eprint {https://arxiv.org/abs/2112.14688} {arXiv:2112.14688 [quant-ph]} \BibitemShut {NoStop}%
\bibitem [{\citenamefont {Mao}\ \emph {et~al.}(2023)\citenamefont {Mao}, \citenamefont {Chaudhary}, \citenamefont {Kondappan}, \citenamefont {Shi}, \citenamefont {Ilo-Okeke}, \citenamefont {Ivannikov},\ and\ \citenamefont {Byrnes}}]{Mao23}%
  \BibitemOpen
  \bibfield  {author} {\bibinfo {author} {\bibfnamefont {Y.}~\bibnamefont {Mao}}, \bibinfo {author} {\bibfnamefont {M.}~\bibnamefont {Chaudhary}}, \bibinfo {author} {\bibfnamefont {M.}~\bibnamefont {Kondappan}}, \bibinfo {author} {\bibfnamefont {J.}~\bibnamefont {Shi}}, \bibinfo {author} {\bibfnamefont {E.~O.}\ \bibnamefont {Ilo-Okeke}}, \bibinfo {author} {\bibfnamefont {V.}~\bibnamefont {Ivannikov}},\ and\ \bibinfo {author} {\bibfnamefont {T.}~\bibnamefont {Byrnes}},\ }\bibfield  {title} {\bibinfo {title} {Measurement-based deterministic imaginary time evolution},\ }\href {https://doi.org/10.1103/PhysRevLett.131.110602} {\bibfield  {journal} {\bibinfo  {journal} {Phys. Rev. Lett.}\ }\textbf {\bibinfo {volume} {131}},\ \bibinfo {pages} {110602} (\bibinfo {year} {2023})}\BibitemShut {NoStop}%
\bibitem [{\citenamefont {Low}\ \emph {et~al.}(2016)\citenamefont {Low}, \citenamefont {Yoder},\ and\ \citenamefont {Chuang}}]{Low16}%
  \BibitemOpen
  \bibfield  {author} {\bibinfo {author} {\bibfnamefont {G.~H.}\ \bibnamefont {Low}}, \bibinfo {author} {\bibfnamefont {T.~J.}\ \bibnamefont {Yoder}},\ and\ \bibinfo {author} {\bibfnamefont {I.~L.}\ \bibnamefont {Chuang}},\ }\bibfield  {title} {\bibinfo {title} {Methodology of resonant equiangular composite quantum gates},\ }\href {https://doi.org/10.1103/PhysRevX.6.041067} {\bibfield  {journal} {\bibinfo  {journal} {Phys. Rev. X}\ }\textbf {\bibinfo {volume} {6}},\ \bibinfo {pages} {041067} (\bibinfo {year} {2016})}\BibitemShut {NoStop}%
\bibitem [{\citenamefont {Low}\ and\ \citenamefont {Chuang}(2017)}]{Low17}%
  \BibitemOpen
  \bibfield  {author} {\bibinfo {author} {\bibfnamefont {G.~H.}\ \bibnamefont {Low}}\ and\ \bibinfo {author} {\bibfnamefont {I.~L.}\ \bibnamefont {Chuang}},\ }\bibfield  {title} {\bibinfo {title} {Optimal hamiltonian simulation by quantum signal processing},\ }\href {https://doi.org/10.1103/PhysRevLett.118.010501} {\bibfield  {journal} {\bibinfo  {journal} {Phys. Rev. Lett.}\ }\textbf {\bibinfo {volume} {118}},\ \bibinfo {pages} {010501} (\bibinfo {year} {2017})}\BibitemShut {NoStop}%
\bibitem [{\citenamefont {Low}\ and\ \citenamefont {Chuang}(2019)}]{Low19}%
  \BibitemOpen
  \bibfield  {author} {\bibinfo {author} {\bibfnamefont {G.~H.}\ \bibnamefont {Low}}\ and\ \bibinfo {author} {\bibfnamefont {I.~L.}\ \bibnamefont {Chuang}},\ }\bibfield  {title} {\bibinfo {title} {Hamiltonian simulation by qubitization},\ }\href {https://doi.org/10.22331/q-2019-07-12-163} {\bibfield  {journal} {\bibinfo  {journal} {Quantum}\ }\textbf {\bibinfo {volume} {3}},\ \bibinfo {pages} {163} (\bibinfo {year} {2019})}\BibitemShut {NoStop}%
\bibitem [{\citenamefont {Martyn}\ \emph {et~al.}(2021)\citenamefont {Martyn}, \citenamefont {Rossi}, \citenamefont {Tan},\ and\ \citenamefont {Chuang}}]{Martyn21}%
  \BibitemOpen
  \bibfield  {author} {\bibinfo {author} {\bibfnamefont {J.~M.}\ \bibnamefont {Martyn}}, \bibinfo {author} {\bibfnamefont {Z.~M.}\ \bibnamefont {Rossi}}, \bibinfo {author} {\bibfnamefont {A.~K.}\ \bibnamefont {Tan}},\ and\ \bibinfo {author} {\bibfnamefont {I.~L.}\ \bibnamefont {Chuang}},\ }\bibfield  {title} {\bibinfo {title} {Grand unification of quantum algorithms},\ }\href {https://doi.org/10.1103/PRXQuantum.2.040203} {\bibfield  {journal} {\bibinfo  {journal} {PRX Quantum}\ }\textbf {\bibinfo {volume} {2}},\ \bibinfo {pages} {040203} (\bibinfo {year} {2021})}\BibitemShut {NoStop}%
\bibitem [{\citenamefont {Coopmans}\ \emph {et~al.}(2023)\citenamefont {Coopmans}, \citenamefont {Kikuchi},\ and\ \citenamefont {Benedetti}}]{Coopmans23}%
  \BibitemOpen
  \bibfield  {author} {\bibinfo {author} {\bibfnamefont {L.}~\bibnamefont {Coopmans}}, \bibinfo {author} {\bibfnamefont {Y.}~\bibnamefont {Kikuchi}},\ and\ \bibinfo {author} {\bibfnamefont {M.}~\bibnamefont {Benedetti}},\ }\bibfield  {title} {\bibinfo {title} {Predicting gibbs-state expectation values with pure thermal shadows},\ }\href {https://doi.org/10.1103/PRXQuantum.4.010305} {\bibfield  {journal} {\bibinfo  {journal} {PRX Quantum}\ }\textbf {\bibinfo {volume} {4}},\ \bibinfo {pages} {010305} (\bibinfo {year} {2023})}\BibitemShut {NoStop}%
\bibitem [{\citenamefont {Poulin}\ and\ \citenamefont {Wocjan}(2009)}]{Poulin09}%
  \BibitemOpen
  \bibfield  {author} {\bibinfo {author} {\bibfnamefont {D.}~\bibnamefont {Poulin}}\ and\ \bibinfo {author} {\bibfnamefont {P.}~\bibnamefont {Wocjan}},\ }\bibfield  {title} {\bibinfo {title} {Sampling from the thermal quantum gibbs state and evaluating partition functions with a quantum computer},\ }\href {https://doi.org/10.1103/PhysRevLett.103.220502} {\bibfield  {journal} {\bibinfo  {journal} {Phys. Rev. Lett.}\ }\textbf {\bibinfo {volume} {103}},\ \bibinfo {pages} {220502} (\bibinfo {year} {2009})}\BibitemShut {NoStop}%
\bibitem [{\citenamefont {Chowdhury}\ and\ \citenamefont {Somma}(2017)}]{Chowdhury17}%
  \BibitemOpen
  \bibfield  {author} {\bibinfo {author} {\bibfnamefont {A.~N.}\ \bibnamefont {Chowdhury}}\ and\ \bibinfo {author} {\bibfnamefont {R.~D.}\ \bibnamefont {Somma}},\ }\bibfield  {title} {\bibinfo {title} {Quantum algorithms for gibbs sampling and hitting-time estimation},\ }\href@noop {} {\bibfield  {journal} {\bibinfo  {journal} {Quantum Info. Comput.}\ }\textbf {\bibinfo {volume} {17}},\ \bibinfo {pages} {41–64} (\bibinfo {year} {2017})}\BibitemShut {NoStop}%
\bibitem [{\citenamefont {Gily\'en}\ \emph {et~al.}(2019)\citenamefont {Gily\'en}, \citenamefont {Su}, \citenamefont {Low},\ and\ \citenamefont {Wiebe}}]{Gilyen19}%
  \BibitemOpen
  \bibfield  {author} {\bibinfo {author} {\bibfnamefont {A.}~\bibnamefont {Gily\'en}}, \bibinfo {author} {\bibfnamefont {Y.}~\bibnamefont {Su}}, \bibinfo {author} {\bibfnamefont {G.~H.}\ \bibnamefont {Low}},\ and\ \bibinfo {author} {\bibfnamefont {N.}~\bibnamefont {Wiebe}},\ }\bibfield  {title} {\bibinfo {title} {Quantum singular value transformation and beyond: exponential improvements for quantum matrix arithmetics},\ }in\ \href {https://doi.org/10.1145/3313276.3316366} {\emph {\bibinfo {booktitle} {Proceedings of the 51st Annual ACM SIGACT Symposium on Theory of Computing}}},\ \bibinfo {series and number} {STOC ’19}\ (\bibinfo  {publisher} {ACM},\ \bibinfo {year} {2019})\BibitemShut {NoStop}%
\bibitem [{\citenamefont {van Apeldoorn}\ \emph {et~al.}(2020)\citenamefont {van Apeldoorn}, \citenamefont {Gily\'en}, \citenamefont {Gribling},\ and\ \citenamefont {de~Wolf}}]{vanApeldoorn20}%
  \BibitemOpen
  \bibfield  {author} {\bibinfo {author} {\bibfnamefont {J.}~\bibnamefont {van Apeldoorn}}, \bibinfo {author} {\bibfnamefont {A.}~\bibnamefont {Gily\'en}}, \bibinfo {author} {\bibfnamefont {S.}~\bibnamefont {Gribling}},\ and\ \bibinfo {author} {\bibfnamefont {R.}~\bibnamefont {de~Wolf}},\ }\bibfield  {title} {\bibinfo {title} {Quantum sdp-solvers: Better upper and lower bounds},\ }\href {https://doi.org/10.22331/q-2020-02-14-230} {\bibfield  {journal} {\bibinfo  {journal} {Quantum}\ }\textbf {\bibinfo {volume} {4}},\ \bibinfo {pages} {230} (\bibinfo {year} {2020})}\BibitemShut {NoStop}%
\bibitem [{\citenamefont {Chowdhury}\ \emph {et~al.}(2021)\citenamefont {Chowdhury}, \citenamefont {Somma},\ and\ \citenamefont {Suba\ifmmode \mbox{\c{s}}\else \c{s}\fi{}\ifmmode \imath \else~\i \fi{}}}]{Chowdhury21}%
  \BibitemOpen
  \bibfield  {author} {\bibinfo {author} {\bibfnamefont {A.~N.}\ \bibnamefont {Chowdhury}}, \bibinfo {author} {\bibfnamefont {R.~D.}\ \bibnamefont {Somma}},\ and\ \bibinfo {author} {\bibfnamefont {Y.~b.~u.}\ \bibnamefont {Suba\ifmmode \mbox{\c{s}}\else \c{s}\fi{}\ifmmode \imath \else~\i \fi{}}},\ }\bibfield  {title} {\bibinfo {title} {Computing partition functions in the one-clean-qubit model},\ }\href {https://doi.org/10.1103/PhysRevA.103.032422} {\bibfield  {journal} {\bibinfo  {journal} {Phys. Rev. A}\ }\textbf {\bibinfo {volume} {103}},\ \bibinfo {pages} {032422} (\bibinfo {year} {2021})}\BibitemShut {NoStop}%
\bibitem [{\citenamefont {Verdon}\ \emph {et~al.}(2019)\citenamefont {Verdon}, \citenamefont {Marks}, \citenamefont {Nanda}, \citenamefont {Leichenauer},\ and\ \citenamefont {Hidary}}]{Verdon19}%
  \BibitemOpen
  \bibfield  {author} {\bibinfo {author} {\bibfnamefont {G.}~\bibnamefont {Verdon}}, \bibinfo {author} {\bibfnamefont {J.}~\bibnamefont {Marks}}, \bibinfo {author} {\bibfnamefont {S.}~\bibnamefont {Nanda}}, \bibinfo {author} {\bibfnamefont {S.}~\bibnamefont {Leichenauer}},\ and\ \bibinfo {author} {\bibfnamefont {J.}~\bibnamefont {Hidary}},\ }\href {https://arxiv.org/abs/1910.02071} {\bibinfo {title} {Quantum hamiltonian-based models and the variational quantum thermalizer algorithm}} (\bibinfo {year} {2019}),\ \Eprint {https://arxiv.org/abs/1910.02071} {arXiv:1910.02071 [quant-ph]} \BibitemShut {NoStop}%
\bibitem [{\citenamefont {Selisko}\ \emph {et~al.}(2023)\citenamefont {Selisko}, \citenamefont {Amsler}, \citenamefont {Hammerschmidt}, \citenamefont {Drautz},\ and\ \citenamefont {Eckl}}]{Selisko23}%
  \BibitemOpen
  \bibfield  {author} {\bibinfo {author} {\bibfnamefont {J.}~\bibnamefont {Selisko}}, \bibinfo {author} {\bibfnamefont {M.}~\bibnamefont {Amsler}}, \bibinfo {author} {\bibfnamefont {T.}~\bibnamefont {Hammerschmidt}}, \bibinfo {author} {\bibfnamefont {R.}~\bibnamefont {Drautz}},\ and\ \bibinfo {author} {\bibfnamefont {T.}~\bibnamefont {Eckl}},\ }\bibfield  {title} {\bibinfo {title} {Extending the variational quantum eigensolver to finite temperatures},\ }\href {https://doi.org/10.1088/2058-9565/ad1340} {\bibfield  {journal} {\bibinfo  {journal} {Quantum Science and Technology}\ }\textbf {\bibinfo {volume} {9}},\ \bibinfo {pages} {015026} (\bibinfo {year} {2023})}\BibitemShut {NoStop}%
\bibitem [{\citenamefont {Pollock}\ \emph {et~al.}(2023)\citenamefont {Pollock}, \citenamefont {Orth},\ and\ \citenamefont {Iadecola}}]{Pollock23}%
  \BibitemOpen
  \bibfield  {author} {\bibinfo {author} {\bibfnamefont {K.}~\bibnamefont {Pollock}}, \bibinfo {author} {\bibfnamefont {P.~P.}\ \bibnamefont {Orth}},\ and\ \bibinfo {author} {\bibfnamefont {T.}~\bibnamefont {Iadecola}},\ }\bibfield  {title} {\bibinfo {title} {Variational microcanonical estimator},\ }\href {https://doi.org/10.1103/PhysRevResearch.5.033224} {\bibfield  {journal} {\bibinfo  {journal} {Phys. Rev. Res.}\ }\textbf {\bibinfo {volume} {5}},\ \bibinfo {pages} {033224} (\bibinfo {year} {2023})}\BibitemShut {NoStop}%
\bibitem [{\citenamefont {Borla}\ \emph {et~al.}(2022)\citenamefont {Borla}, \citenamefont {Jeevanesan}, \citenamefont {Pollmann},\ and\ \citenamefont {Moroz}}]{Borla22}%
  \BibitemOpen
  \bibfield  {author} {\bibinfo {author} {\bibfnamefont {U.}~\bibnamefont {Borla}}, \bibinfo {author} {\bibfnamefont {B.}~\bibnamefont {Jeevanesan}}, \bibinfo {author} {\bibfnamefont {F.}~\bibnamefont {Pollmann}},\ and\ \bibinfo {author} {\bibfnamefont {S.}~\bibnamefont {Moroz}},\ }\bibfield  {title} {\bibinfo {title} {Quantum phases of two-dimensional ${Z}_{2}$ gauge theory coupled to single-component fermion matter},\ }\href {https://doi.org/10.1103/PhysRevB.105.075132} {\bibfield  {journal} {\bibinfo  {journal} {Phys. Rev. B}\ }\textbf {\bibinfo {volume} {105}},\ \bibinfo {pages} {075132} (\bibinfo {year} {2022})}\BibitemShut {NoStop}%
\bibitem [{\citenamefont {Celi}\ \emph {et~al.}(2020)\citenamefont {Celi}, \citenamefont {Vermersch}, \citenamefont {Viyuela}, \citenamefont {Pichler}, \citenamefont {Lukin},\ and\ \citenamefont {Zoller}}]{Celi20}%
  \BibitemOpen
  \bibfield  {author} {\bibinfo {author} {\bibfnamefont {A.}~\bibnamefont {Celi}}, \bibinfo {author} {\bibfnamefont {B.}~\bibnamefont {Vermersch}}, \bibinfo {author} {\bibfnamefont {O.}~\bibnamefont {Viyuela}}, \bibinfo {author} {\bibfnamefont {H.}~\bibnamefont {Pichler}}, \bibinfo {author} {\bibfnamefont {M.~D.}\ \bibnamefont {Lukin}},\ and\ \bibinfo {author} {\bibfnamefont {P.}~\bibnamefont {Zoller}},\ }\bibfield  {title} {\bibinfo {title} {Emerging two-dimensional gauge theories in {Rydberg} configurable arrays},\ }\bibfield  {journal} {\bibinfo  {journal} {Phys.~Rev.~X}\ }\textbf {\bibinfo {volume} {10}},\ \href {https://doi.org/10.1103/physrevx.10.021057} {10.1103/physrevx.10.021057} (\bibinfo {year} {2020})\BibitemShut {NoStop}%
\bibitem [{\citenamefont {Chen}\ and\ \citenamefont {Iadecola}(2021)}]{Chen21}%
  \BibitemOpen
  \bibfield  {author} {\bibinfo {author} {\bibfnamefont {I.-C.}\ \bibnamefont {Chen}}\ and\ \bibinfo {author} {\bibfnamefont {T.}~\bibnamefont {Iadecola}},\ }\bibfield  {title} {\bibinfo {title} {Emergent symmetries and slow quantum dynamics in a rydberg-atom chain with confinement},\ }\href {https://doi.org/10.1103/PhysRevB.103.214304} {\bibfield  {journal} {\bibinfo  {journal} {Phys. Rev. B}\ }\textbf {\bibinfo {volume} {103}},\ \bibinfo {pages} {214304} (\bibinfo {year} {2021})}\BibitemShut {NoStop}%
\bibitem [{\citenamefont {Silvi}\ \emph {et~al.}(2019)\citenamefont {Silvi}, \citenamefont {Sauer}, \citenamefont {Tschirsich},\ and\ \citenamefont {Montangero}}]{Silvi19}%
  \BibitemOpen
  \bibfield  {author} {\bibinfo {author} {\bibfnamefont {P.}~\bibnamefont {Silvi}}, \bibinfo {author} {\bibfnamefont {Y.}~\bibnamefont {Sauer}}, \bibinfo {author} {\bibfnamefont {F.}~\bibnamefont {Tschirsich}},\ and\ \bibinfo {author} {\bibfnamefont {S.}~\bibnamefont {Montangero}},\ }\bibfield  {title} {\bibinfo {title} {Tensor network simulation of an su(3) lattice gauge theory in 1d},\ }\href {https://doi.org/10.1103/PhysRevD.100.074512} {\bibfield  {journal} {\bibinfo  {journal} {Phys. Rev. D}\ }\textbf {\bibinfo {volume} {100}},\ \bibinfo {pages} {074512} (\bibinfo {year} {2019})}\BibitemShut {NoStop}%
\bibitem [{\citenamefont {Rigobello}\ \emph {et~al.}(2023)\citenamefont {Rigobello}, \citenamefont {Magnifico}, \citenamefont {Silvi},\ and\ \citenamefont {Montangero}}]{Rigobello23}%
  \BibitemOpen
  \bibfield  {author} {\bibinfo {author} {\bibfnamefont {M.}~\bibnamefont {Rigobello}}, \bibinfo {author} {\bibfnamefont {G.}~\bibnamefont {Magnifico}}, \bibinfo {author} {\bibfnamefont {P.}~\bibnamefont {Silvi}},\ and\ \bibinfo {author} {\bibfnamefont {S.}~\bibnamefont {Montangero}},\ }\href@noop {} {\bibinfo {title} {Hadrons in (1+1)d hamiltonian hardcore lattice qcd}} (\bibinfo {year} {2023}),\ \Eprint {https://arxiv.org/abs/arXiv:2308.04488} {arXiv:2308.04488} \BibitemShut {NoStop}%
\bibitem [{\citenamefont {Cataldi}\ \emph {et~al.}(2024)\citenamefont {Cataldi}, \citenamefont {Magnifico}, \citenamefont {Silvi},\ and\ \citenamefont {Montangero}}]{Cataldi24}%
  \BibitemOpen
  \bibfield  {author} {\bibinfo {author} {\bibfnamefont {G.}~\bibnamefont {Cataldi}}, \bibinfo {author} {\bibfnamefont {G.}~\bibnamefont {Magnifico}}, \bibinfo {author} {\bibfnamefont {P.}~\bibnamefont {Silvi}},\ and\ \bibinfo {author} {\bibfnamefont {S.}~\bibnamefont {Montangero}},\ }\bibfield  {title} {\bibinfo {title} {Simulating $(2+1)\mathrm{D}$ su(2) yang-mills lattice gauge theory at finite density with tensor networks},\ }\href {https://doi.org/10.1103/PhysRevResearch.6.033057} {\bibfield  {journal} {\bibinfo  {journal} {Phys. Rev. Res.}\ }\textbf {\bibinfo {volume} {6}},\ \bibinfo {pages} {033057} (\bibinfo {year} {2024})}\BibitemShut {NoStop}%
\bibitem [{\citenamefont {Calaj\'o}\ \emph {et~al.}(2024)\citenamefont {Calaj\'o}, \citenamefont {Magnifico}, \citenamefont {Edmunds}, \citenamefont {Ringbauer}, \citenamefont {Montangero},\ and\ \citenamefont {Silvi}}]{Calajo24}%
  \BibitemOpen
  \bibfield  {author} {\bibinfo {author} {\bibfnamefont {G.}~\bibnamefont {Calaj\'o}}, \bibinfo {author} {\bibfnamefont {G.}~\bibnamefont {Magnifico}}, \bibinfo {author} {\bibfnamefont {C.}~\bibnamefont {Edmunds}}, \bibinfo {author} {\bibfnamefont {M.}~\bibnamefont {Ringbauer}}, \bibinfo {author} {\bibfnamefont {S.}~\bibnamefont {Montangero}},\ and\ \bibinfo {author} {\bibfnamefont {P.}~\bibnamefont {Silvi}},\ }\bibfield  {title} {\bibinfo {title} {Digital quantum simulation of a (1+1)d su(2) lattice gauge theory with ion qudits},\ }\href {https://doi.org/10.1103/PRXQuantum.5.040309} {\bibfield  {journal} {\bibinfo  {journal} {PRX Quantum}\ }\textbf {\bibinfo {volume} {5}},\ \bibinfo {pages} {040309} (\bibinfo {year} {2024})}\BibitemShut {NoStop}%
\bibitem [{\citenamefont {Zohar}\ and\ \citenamefont {Cirac}(2018)}]{Zohar18}%
  \BibitemOpen
  \bibfield  {author} {\bibinfo {author} {\bibfnamefont {E.}~\bibnamefont {Zohar}}\ and\ \bibinfo {author} {\bibfnamefont {J.~I.}\ \bibnamefont {Cirac}},\ }\bibfield  {title} {\bibinfo {title} {Eliminating fermionic matter fields in lattice gauge theories},\ }\href {https://doi.org/10.1103/PhysRevB.98.075119} {\bibfield  {journal} {\bibinfo  {journal} {Phys. Rev. B}\ }\textbf {\bibinfo {volume} {98}},\ \bibinfo {pages} {075119} (\bibinfo {year} {2018})}\BibitemShut {NoStop}%
\bibitem [{\citenamefont {Watts}\ \emph {et~al.}(2016)\citenamefont {Watts} \emph {et~al.}}]{Watts:2016uzu}%
  \BibitemOpen
  \bibfield  {author} {\bibinfo {author} {\bibfnamefont {A.~L.}\ \bibnamefont {Watts}} \emph {et~al.},\ }\bibfield  {title} {\bibinfo {title} {{Colloquium : Measuring the neutron star equation of state using x-ray timing}},\ }\href {https://doi.org/10.1103/RevModPhys.88.021001} {\bibfield  {journal} {\bibinfo  {journal} {Rev. Mod. Phys.}\ }\textbf {\bibinfo {volume} {88}},\ \bibinfo {pages} {021001} (\bibinfo {year} {2016})},\ \Eprint {https://arxiv.org/abs/1602.01081} {arXiv:1602.01081 [astro-ph.HE]} \BibitemShut {NoStop}%
\bibitem [{\citenamefont {\"Ozel}\ and\ \citenamefont {Freire}(2016)}]{Ozel:2016oaf}%
  \BibitemOpen
  \bibfield  {author} {\bibinfo {author} {\bibfnamefont {F.}~\bibnamefont {\"Ozel}}\ and\ \bibinfo {author} {\bibfnamefont {P.}~\bibnamefont {Freire}},\ }\bibfield  {title} {\bibinfo {title} {{Masses, Radii, and the Equation of State of Neutron Stars}},\ }\href {https://doi.org/10.1146/annurev-astro-081915-023322} {\bibfield  {journal} {\bibinfo  {journal} {Ann. Rev. Astron. Astrophys.}\ }\textbf {\bibinfo {volume} {54}},\ \bibinfo {pages} {401} (\bibinfo {year} {2016})},\ \Eprint {https://arxiv.org/abs/1603.02698} {arXiv:1603.02698 [astro-ph.HE]} \BibitemShut {NoStop}%
\bibitem [{\citenamefont {Demorest}\ \emph {et~al.}(2010)\citenamefont {Demorest}, \citenamefont {Pennucci}, \citenamefont {Ransom}, \citenamefont {Roberts},\ and\ \citenamefont {Hessels}}]{Demorest:2010bx}%
  \BibitemOpen
  \bibfield  {author} {\bibinfo {author} {\bibfnamefont {P.}~\bibnamefont {Demorest}}, \bibinfo {author} {\bibfnamefont {T.}~\bibnamefont {Pennucci}}, \bibinfo {author} {\bibfnamefont {S.}~\bibnamefont {Ransom}}, \bibinfo {author} {\bibfnamefont {M.}~\bibnamefont {Roberts}},\ and\ \bibinfo {author} {\bibfnamefont {J.}~\bibnamefont {Hessels}},\ }\bibfield  {title} {\bibinfo {title} {{Shapiro Delay Measurement of A Two Solar Mass Neutron Star}},\ }\href {https://doi.org/10.1038/nature09466} {\bibfield  {journal} {\bibinfo  {journal} {Nature}\ }\textbf {\bibinfo {volume} {467}},\ \bibinfo {pages} {1081} (\bibinfo {year} {2010})},\ \Eprint {https://arxiv.org/abs/1010.5788} {arXiv:1010.5788 [astro-ph.HE]} \BibitemShut {NoStop}%
\bibitem [{\citenamefont {Antoniadis}\ \emph {et~al.}(2013)\citenamefont {Antoniadis} \emph {et~al.}}]{Antoniadis:2013pzd}%
  \BibitemOpen
  \bibfield  {author} {\bibinfo {author} {\bibfnamefont {J.}~\bibnamefont {Antoniadis}} \emph {et~al.},\ }\bibfield  {title} {\bibinfo {title} {{A Massive Pulsar in a Compact Relativistic Binary}},\ }\href {https://doi.org/10.1126/science.1233232} {\bibfield  {journal} {\bibinfo  {journal} {Science}\ }\textbf {\bibinfo {volume} {340}},\ \bibinfo {pages} {6131} (\bibinfo {year} {2013})},\ \Eprint {https://arxiv.org/abs/1304.6875} {arXiv:1304.6875 [astro-ph.HE]} \BibitemShut {NoStop}%
\bibitem [{\citenamefont {Radice}\ \emph {et~al.}(2018)\citenamefont {Radice}, \citenamefont {Perego}, \citenamefont {Zappa},\ and\ \citenamefont {Bernuzzi}}]{Radice:2017lry}%
  \BibitemOpen
  \bibfield  {author} {\bibinfo {author} {\bibfnamefont {D.}~\bibnamefont {Radice}}, \bibinfo {author} {\bibfnamefont {A.}~\bibnamefont {Perego}}, \bibinfo {author} {\bibfnamefont {F.}~\bibnamefont {Zappa}},\ and\ \bibinfo {author} {\bibfnamefont {S.}~\bibnamefont {Bernuzzi}},\ }\bibfield  {title} {\bibinfo {title} {{GW170817: Joint Constraint on the Neutron Star Equation of State from Multimessenger Observations}},\ }\href {https://doi.org/10.3847/2041-8213/aaa402} {\bibfield  {journal} {\bibinfo  {journal} {Astrophys. J. Lett.}\ }\textbf {\bibinfo {volume} {852}},\ \bibinfo {pages} {L29} (\bibinfo {year} {2018})},\ \Eprint {https://arxiv.org/abs/1711.03647} {arXiv:1711.03647 [astro-ph.HE]} \BibitemShut {NoStop}%
\bibitem [{\citenamefont {Bauswein}\ and\ \citenamefont {Janka}(2012)}]{PhysRevLett.108.011101}%
  \BibitemOpen
  \bibfield  {author} {\bibinfo {author} {\bibfnamefont {A.}~\bibnamefont {Bauswein}}\ and\ \bibinfo {author} {\bibfnamefont {H.-T.}\ \bibnamefont {Janka}},\ }\bibfield  {title} {\bibinfo {title} {Measuring neutron-star properties via gravitational waves from neutron-star mergers},\ }\href {https://doi.org/10.1103/PhysRevLett.108.011101} {\bibfield  {journal} {\bibinfo  {journal} {Phys. Rev. Lett.}\ }\textbf {\bibinfo {volume} {108}},\ \bibinfo {pages} {011101} (\bibinfo {year} {2012})}\BibitemShut {NoStop}%
\bibitem [{\citenamefont {Takami}\ \emph {et~al.}(2014)\citenamefont {Takami}, \citenamefont {Rezzolla},\ and\ \citenamefont {Baiotti}}]{Takami:2014zpa}%
  \BibitemOpen
  \bibfield  {author} {\bibinfo {author} {\bibfnamefont {K.}~\bibnamefont {Takami}}, \bibinfo {author} {\bibfnamefont {L.}~\bibnamefont {Rezzolla}},\ and\ \bibinfo {author} {\bibfnamefont {L.}~\bibnamefont {Baiotti}},\ }\bibfield  {title} {\bibinfo {title} {{Constraining the Equation of State of Neutron Stars from Binary Mergers}},\ }\href {https://doi.org/10.1103/PhysRevLett.113.091104} {\bibfield  {journal} {\bibinfo  {journal} {Phys. Rev. Lett.}\ }\textbf {\bibinfo {volume} {113}},\ \bibinfo {pages} {091104} (\bibinfo {year} {2014})},\ \Eprint {https://arxiv.org/abs/1403.5672} {arXiv:1403.5672 [gr-qc]} \BibitemShut {NoStop}%
\bibitem [{\citenamefont {Radice}\ \emph {et~al.}(2017)\citenamefont {Radice}, \citenamefont {Bernuzzi}, \citenamefont {Del~Pozzo}, \citenamefont {Roberts},\ and\ \citenamefont {Ott}}]{Radice:2016rys}%
  \BibitemOpen
  \bibfield  {author} {\bibinfo {author} {\bibfnamefont {D.}~\bibnamefont {Radice}}, \bibinfo {author} {\bibfnamefont {S.}~\bibnamefont {Bernuzzi}}, \bibinfo {author} {\bibfnamefont {W.}~\bibnamefont {Del~Pozzo}}, \bibinfo {author} {\bibfnamefont {L.~F.}\ \bibnamefont {Roberts}},\ and\ \bibinfo {author} {\bibfnamefont {C.~D.}\ \bibnamefont {Ott}},\ }\bibfield  {title} {\bibinfo {title} {{Probing Extreme-Density Matter with Gravitational Wave Observations of Binary Neutron Star Merger Remnants}},\ }\href {https://doi.org/10.3847/2041-8213/aa775f} {\bibfield  {journal} {\bibinfo  {journal} {Astrophys. J. Lett.}\ }\textbf {\bibinfo {volume} {842}},\ \bibinfo {pages} {L10} (\bibinfo {year} {2017})},\ \Eprint {https://arxiv.org/abs/1612.06429} {arXiv:1612.06429 [astro-ph.HE]} \BibitemShut {NoStop}%
\bibitem [{\citenamefont {Chatziioannou}\ \emph {et~al.}(2017)\citenamefont {Chatziioannou}, \citenamefont {Clark}, \citenamefont {Bauswein}, \citenamefont {Millhouse}, \citenamefont {Littenberg},\ and\ \citenamefont {Cornish}}]{Chatziioannou:2017ixj}%
  \BibitemOpen
  \bibfield  {author} {\bibinfo {author} {\bibfnamefont {K.}~\bibnamefont {Chatziioannou}}, \bibinfo {author} {\bibfnamefont {J.~A.}\ \bibnamefont {Clark}}, \bibinfo {author} {\bibfnamefont {A.}~\bibnamefont {Bauswein}}, \bibinfo {author} {\bibfnamefont {M.}~\bibnamefont {Millhouse}}, \bibinfo {author} {\bibfnamefont {T.~B.}\ \bibnamefont {Littenberg}},\ and\ \bibinfo {author} {\bibfnamefont {N.}~\bibnamefont {Cornish}},\ }\bibfield  {title} {\bibinfo {title} {{Inferring the post-merger gravitational wave emission from binary neutron star coalescences}},\ }\href {https://doi.org/10.1103/PhysRevD.96.124035} {\bibfield  {journal} {\bibinfo  {journal} {Phys. Rev. D}\ }\textbf {\bibinfo {volume} {96}},\ \bibinfo {pages} {124035} (\bibinfo {year} {2017})},\ \Eprint {https://arxiv.org/abs/1711.00040} {arXiv:1711.00040 [gr-qc]} \BibitemShut {NoStop}%
\bibitem [{\citenamefont {Bedaque}\ and\ \citenamefont {Steiner}(2015)}]{Bedaque:2014sqa}%
  \BibitemOpen
  \bibfield  {author} {\bibinfo {author} {\bibfnamefont {P.}~\bibnamefont {Bedaque}}\ and\ \bibinfo {author} {\bibfnamefont {A.~W.}\ \bibnamefont {Steiner}},\ }\bibfield  {title} {\bibinfo {title} {{Sound velocity bound and neutron stars}},\ }\href {https://doi.org/10.1103/PhysRevLett.114.031103} {\bibfield  {journal} {\bibinfo  {journal} {Phys. Rev. Lett.}\ }\textbf {\bibinfo {volume} {114}},\ \bibinfo {pages} {031103} (\bibinfo {year} {2015})},\ \Eprint {https://arxiv.org/abs/1408.5116} {arXiv:1408.5116 [nucl-th]} \BibitemShut {NoStop}%
\end{thebibliography}%

\end{document}